\documentclass[11pt,a4paper]{article} 
\usepackage{graphicx}
\usepackage{style}
\usepackage{epstopdf}
\usepackage{epsfig}
\usepackage{amsmath,amssymb,bm,amsfonts,yfonts}
\newcommand{\be}{\begin{equation}}
\newcommand{\ee}{\end{equation}}
\newcommand{\bea}{\begin{eqnarray}}
\newcommand{\nn}{\nonumber}
\newcommand{\eea}{\end{eqnarray}}

\def\Hc{{\cal H}}
\newcommand{\alphat}{\tilde{\alpha}}
\newcommand{\phit}{\tilde{\phi}}
\newcommand{\etat}{{\tilde{\eta}}}

\newcommand{\td}{{\delta}}

\newcommand{\drho}{\delta \rho}
\newcommand{\rhoz}{\rho_m}
\newcommand{\pz}{p_m}
\newcommand{\dpp}{\delta p}

\newcommand{\bp}{\mathbf{p}}
\newcommand{\bq}{\mathbf{q}}
\newcommand{\bk}{\mathbf{k}}

\newcommand{\bx}{\mathbf{x}}

\newcommand{\Omegat}{{\tilde{\Omega}}}

\DeclareSymbolFont{mathscrUC}{U}{rsfs}{m}{n}  
\DeclareSymbolFont{mathscrLC}{OT1}{pzc}{m}{n} 

\makeatletter

\DeclareRobustCommand*{\mathscr}[1]{\gdef\F@ntPrefix{mathscr@char@}%
  \@EachCharacter #1\@EndEachCharacter}
\long\def\DoLongFutureLet #1#2#3#4{%
   \def\@FutureLetDecide{#1#2\@FutureLetToken
      \def\@FutureLetNext{#3}\else
      \def\@FutureLetNext{#4}\fi\@FutureLetNext}
   \futurelet\@FutureLetToken\@FutureLetDecide}
\def\DoFutureLet #1#2#3#4{\DoLongFutureLet{#1}{#2}{#3}{#4}}
\def\@EachCharacter{\DoFutureLet{\ifx}{\@EndEachCharacter}%
   {\@EachCharacterDone}{\@PickUpTheCharacter}}
\def\m@keCharacter#1{\csname\F@ntPrefix#1\endcsname}
\def\@PickUpTheCharacter#1{\m@keCharacter{#1}\@EachCharacter}
\def\@EachCharacterDone \@EndEachCharacter{}

\DeclareMathSymbol{\mathscr@char@A}{\mathord}{mathscrUC}{`A}
\DeclareMathSymbol{\mathscr@char@B}{\mathord}{mathscrUC}{`B}
\DeclareMathSymbol{\mathscr@char@C}{\mathord}{mathscrUC}{`C}
\DeclareMathSymbol{\mathscr@char@D}{\mathord}{mathscrUC}{`D}
\DeclareMathSymbol{\mathscr@char@E}{\mathord}{mathscrUC}{`E}
\DeclareMathSymbol{\mathscr@char@F}{\mathord}{mathscrUC}{`F}
\DeclareMathSymbol{\mathscr@char@G}{\mathord}{mathscrUC}{`G}
\DeclareMathSymbol{\mathscr@char@H}{\mathord}{mathscrUC}{`H}
\DeclareMathSymbol{\mathscr@char@I}{\mathord}{mathscrUC}{`I}
\DeclareMathSymbol{\mathscr@char@J}{\mathord}{mathscrUC}{`J}
\DeclareMathSymbol{\mathscr@char@K}{\mathord}{mathscrUC}{`K}
\DeclareMathSymbol{\mathscr@char@L}{\mathord}{mathscrUC}{`L}
\DeclareMathSymbol{\mathscr@char@M}{\mathord}{mathscrUC}{`M}
\DeclareMathSymbol{\mathscr@char@N}{\mathord}{mathscrUC}{`N}
\DeclareMathSymbol{\mathscr@char@O}{\mathord}{mathscrUC}{`O}
\DeclareMathSymbol{\mathscr@char@P}{\mathord}{mathscrUC}{`P}
\DeclareMathSymbol{\mathscr@char@Q}{\mathord}{mathscrUC}{`Q}
\DeclareMathSymbol{\mathscr@char@R}{\mathord}{mathscrUC}{`R}
\DeclareMathSymbol{\mathscr@char@S}{\mathord}{mathscrUC}{`S}
\DeclareMathSymbol{\mathscr@char@T}{\mathord}{mathscrUC}{`T}
\DeclareMathSymbol{\mathscr@char@U}{\mathord}{mathscrUC}{`U}
\DeclareMathSymbol{\mathscr@char@V}{\mathord}{mathscrUC}{`V}
\DeclareMathSymbol{\mathscr@char@W}{\mathord}{mathscrUC}{`W}
\DeclareMathSymbol{\mathscr@char@X}{\mathord}{mathscrUC}{`X}
\DeclareMathSymbol{\mathscr@char@Y}{\mathord}{mathscrUC}{`Y}
\DeclareMathSymbol{\mathscr@char@Z}{\mathord}{mathscrUC}{`Z}
\DeclareMathSymbol{\mathscr@char@a}{\mathord}{mathscrLC}{`a}
\DeclareMathSymbol{\mathscr@char@b}{\mathord}{mathscrLC}{`b}
\DeclareMathSymbol{\mathscr@char@c}{\mathord}{mathscrLC}{`c}
\DeclareMathSymbol{\mathscr@char@d}{\mathord}{mathscrLC}{`d}
\DeclareMathSymbol{\mathscr@char@e}{\mathord}{mathscrLC}{`e}
\DeclareMathSymbol{\mathscr@char@f}{\mathord}{mathscrLC}{`f}
\DeclareMathSymbol{\mathscr@char@g}{\mathord}{mathscrLC}{`g}
\DeclareMathSymbol{\mathscr@char@h}{\mathord}{mathscrLC}{`h}
\DeclareMathSymbol{\mathscr@char@i}{\mathord}{mathscrLC}{`i}
\DeclareMathSymbol{\mathscr@char@j}{\mathord}{mathscrLC}{`j}
\DeclareMathSymbol{\mathscr@char@k}{\mathord}{mathscrLC}{`k}
\DeclareMathSymbol{\mathscr@char@l}{\mathord}{mathscrLC}{`l}
\DeclareMathSymbol{\mathscr@char@m}{\mathord}{mathscrLC}{`m}
\DeclareMathSymbol{\mathscr@char@n}{\mathord}{mathscrLC}{`n}
\DeclareMathSymbol{\mathscr@char@o}{\mathord}{mathscrLC}{`o}
\DeclareMathSymbol{\mathscr@char@p}{\mathord}{mathscrLC}{`p}
\DeclareMathSymbol{\mathscr@char@q}{\mathord}{mathscrLC}{`q}
\DeclareMathSymbol{\mathscr@char@r}{\mathord}{mathscrLC}{`r}
\DeclareMathSymbol{\mathscr@char@s}{\mathord}{mathscrLC}{`s}
\DeclareMathSymbol{\mathscr@char@t}{\mathord}{mathscrLC}{`t}
\DeclareMathSymbol{\mathscr@char@u}{\mathord}{mathscrLC}{`u}
\DeclareMathSymbol{\mathscr@char@v}{\mathord}{mathscrLC}{`v}
\DeclareMathSymbol{\mathscr@char@w}{\mathord}{mathscrLC}{`w}
\DeclareMathSymbol{\mathscr@char@x}{\mathord}{mathscrLC}{`x}
\DeclareMathSymbol{\mathscr@char@y}{\mathord}{mathscrLC}{`y}
\DeclareMathSymbol{\mathscr@char@z}{\mathord}{mathscrLC}{`z}

\makeatother

\relax

\title{Large scale structure from viscous dark matter}

\author[a]{Diego Blas,} 
\author[a]{Stefan Floerchinger,}
\author[a]{Mathias Garny,}
\author[a,b]{Nikolaos Tetradis,}
\author[a]{Urs Achim Wiedemann}

\affiliation[a]{Physics Department, Theory Unit, CERN, CH-1211 Gen\`eve 23, Switzerland}
\affiliation[b]{Department of Physics, University of Athens, Zographou 157 84, Greece}

\emailAdd{diego.blas@cern.ch}
\emailAdd{stefan.floerchinger@cern.ch}
\emailAdd{mathias.garny@cern.ch}
\emailAdd{ntetrad@phys.uoa.gr}
\emailAdd{urs.wiedemann@cern.ch}

\abstract{Cosmological perturbations of sufficiently long wavelength admit a fluid dynamic description. We consider modes with wavevectors below a scale $k_m$ for which the dynamics is only mildly non-linear. The leading effect of modes above that scale can be accounted for by effective non-equilibrium viscosity and pressure terms. For mildly non-linear scales, these mainly arise from momentum transport within the ideal and cold but inhomogeneous fluid, while momentum transport due to more microscopic degrees of freedom is suppressed. As a consequence, concrete expressions with no free parameters, except the matching scale $k_m$, can be derived from matching evolution equations to standard cosmological perturbation theory.
Two-loop calculations of the matter power spectrum in the viscous theory lead to excellent agreement with $N$-body simulations up to scales $k=0.2 \, h/$Mpc. The convergence properties in the ultraviolet are better than for standard perturbation theory and the results are robust with respect to variations of the matching scale.}

\begin{document}

\maketitle

\section{Introduction} \label{intro}

The evolution of cosmological perturbations and their growth at late times is a problem of fundamental significance in cosmology. 
The small amplitude of the primordial density perturbations over the isotropic and homogenous background makes
linear perturbation theory a good description at early times.  
However, at late times the gravitational instability leads to growth of perturbations, inducing the emergence of large-scale structure
and the formation of galaxies and galaxy clusters. As a result, linear perturbation theory ceases to be an accurate framework. 
Going beyond a linearized approach becomes mandatory for a precise comparison of theoretical predictions and astrophysical
observations on length scales of ${\cal O}(100\, {\rm Mpc})$ and below, where density and velocity perturbations are sufficiently large
to render the evolution non-linear but still amenable to perturbative approaches.
This range of scales is interesting because it displays the imprint of the oscillations of the primordial plasma on the matter density field,
the so-called baryon acoustic oscillations (BAO). 
In this work, we show that viscous fluid dynamics provides a simple and accurate framework for extending the description
of cosmological perturbations into this (mildly) non-linear regime. Compared to other formulations that we review below, our approach
is novel in that it fixes the (effective) viscous coefficients and the pressure from a perturbative calculation and without
invoking any fit to non-perturbative information such as results from $N$-body simulations. 

The main analytical approach that has been developed to go beyond linear theory aims at accounting
for interactions between cosmological perturbations of different wavelength perturbatively in the amplitude of the primordial perturbations \cite{Juszkiewicz,Vishniac,Fry:1983cj,Goroff:1986ep}, see ~\cite{bernardeau0,bernardeau} for reviews. This standard perturbation theory (SPT) is successful in producing the first non-linear corrections, but it is not a systematic approach for accurate calculations 
even for mildly non-linear scales. In particular, its convergence at the two-loop level becomes problematic at scales around 30-50 Mpc  and 
the  three-loop contribution becomes comparable to the linear 
spectrum at even larger length scales \cite{Blas1}. The physical origin of the apparent lack of convergence is that higher orders  in the perturbative expansion become more and more sensitive to ultraviolet (UV) modes whose perturbative treatment is questionable. This has triggered the development
of approaches beyond SPT aiming at a better control of perturbation theory at scales up
 to 10 Mpc \cite{CrSc1,CrSc2,taruya,taruya2,Max1,time,mcdonald,Blas1,Wang,Crocce:2007dt,McDonald:2009hs}. 
Gaining a quantitative understanding of the BAO range motivates also the present work. 

One recent approach that aims at addressing this issue borrows ideas from the framework of effective field theory~\cite{Baumann:2010tm,Carrasco:2012cv,eff1,eff2,eff3,eff4,Manzotti,Porto:2013qua,Baldauf:2015aha}. 
The suggestion is to treat the equations describing the large-scale perturbations as reflecting effective properties of dark matter that 
arise from integrating out the short-distance modes up to some coarse-graining scale. The implementation of this essentially Wilsonian point of 
view is complicated by the explicit time dependence of the process of growth of perturbations, and the resulting effective theory in 
general is non-local in time. Despite these difficulties, the formalism has been developed at the two-loop level and it compares well
to the matter power spectrum of simulations down to scales of around 10 Mpc \cite{eff4,Baldauf:2015aha}. The counterterms in this effective field
theory of cosmological perturbations have the purpose of removing the dependence of loop corrections on the UV cutoff. They enter
the effective energy-momentum tensor in the form of an effective pressure and viscosity coefficients, and their residual finite contribution
to the spectrum is fixed by comparing with $N$-body simulations. 

Here, we follow a complementary approach to the growth of cosmological perturbations by {\it starting} from a viscous fluid dynamic formulation.
This approach is motivated by the fact that essentially all physical systems lend themselves to a (not necessarily ideal) fluid dynamic description on sufficiently long time scales and for sufficiently long wavelengths \cite{Hohenberg:1977ym}.
Viscous fluid dynamics is an extension of ideal fluid dynamics that accounts for deviations from this idealization in a gradient expansion. 
As we shall discuss in detail in the following, the viscous fluid dynamic approach to large scale structure has some commonalities with the 
effective field theory approach discussed above. In the present work we focus on the physical origin of the non-ideal behavior on large scales. For this, it is conceptually important to realize that applications of viscous fluid dynamics allow for viscosities of two qualitatively different origins: 
\begin{itemize}
\item \emph{Fundamental (microscopic) viscosity:}
This viscosity is due to the diffusive transport of momentum by the microscopic constituents of the fluid, that is, by particles or radiation. 
It is a fundamental property of the matter under consideration that can be traced back to the fundamental interactions between the fluid
constituents. In cases in which the quantum field theory of the matter is known, these fundamental viscosities are calculable from the
Lagrangian via the Green-Kubo formula \cite{Green,Kubo}.
\item \emph{Effective viscosity:} 
If one limits a fluid dynamic description to large wavelengths, i.e. wavenumbers $k$ below a certain matching scale $k_m$, then dissipation of these fluid perturbations
can be mediated by fluid dynamic degrees of freedom on scales $k > k_m$. The use of effective viscosities parametrizes then how
momentum carried by the long wavelength modes dissipates to shorter wavelengths ($k> k_m$) that are fluid dynamic in nature but that
are not followed in the large wavelength fluid dynamic description.\footnote{The fluid description is expected to break down on very short length scales. The corresponding modes, in principle, can affect the long scale dynamics. However, as discussed in section \ref{sec3}, there is evidence that this effect is small on BAO scales.}
The physical mechanism underlying this is the interaction between long and short wavelengths 
modes. Examples are ubiquitous and include e.g. the description of approximately laminar flow on large scales in the presence of short scale 
eddies (so-called eddy viscosity) \cite{Frisch}. The introduction of effective viscosities is a method of choice, in 
particular if one aims at limiting an application of fluid dynamics to scales that evolve with only mildly non-linear effects (where semi-analytical
methods apply), while smaller length scales show a strong non-linear (even turbulent) behavior with certain degrees of freedom even outside of 
the fluid approximation. In contrast to fundamental viscosity, 
effective viscosity is not a fundamental property of the matter, but it depends on the spectrum of excitations $k>k_m$ with which the
long-wavelength modes can interact, and on the matching scale $k_m$ up to which these long-wavelength modes are evolved explicitly in the
fluid dynamic description.\footnote{Microscopic viscosity could, however, play a role at short wavelength. There could also be an interplay between the effective and the microscopic viscosity, in the sense that momentum is transported to the short wavelength fluid modes by the former, from where it is then dissipated by the latter.}
\end{itemize}

We emphasize that the viscous fluid dynamic formalism introduced in section~\ref{sec2} allows 
for exploring observable consequences of dark matter with either fundamental or effective viscosities and pressure.
The fundamental (microscopic) material properties of the cosmological fluid at late times (dominated by dark matter) may have observational consequences on galaxy cluster scales \cite{Rocha:2012jg,Kahlhoefer:2013dca,Cyr-Racine:2013fsa,Massey:2015dkw,Kahlhoefer:2015vua,Randall:2014kta}, as well as on very large cosmological scales \cite{Maartens:1995wt,Zimdahl:1996ka,Fabris:2005ts,Li:2009mf,Gagnon:2011id,Velten:2012uv,Velten:2013pra,Floerchinger:2014jsa}. Still it is well conceivable that microscopic viscosity and pressure are negligible for the evolution of cosmological perturbations up to the BAO range.  In particular, they may be much smaller than
 the effective viscosity and pressure of a fluid dynamic description limited to $k < k_m \sim {\cal O}\left(1\, h/\text{Mpc} \right)$. An analogous statement holds for the notion of pressure: dark matter may be essentially pressureless on a fundamental level, while an effective
pressure can arise if fluid dynamics is limited to $k < k_m$. 

In the present work, we explore specifically the possibility that effective viscosities and pressure dominate for the description of scales $k < k_m \sim {\cal O}\left( 1\, h\,/{\rm  Mpc} \right)$. 
For this case, we argue in section~\ref{sec3} that realistic numerical proxies for the effective viscosities and pressure can be obtained by matching
viscous fluid dynamics to results from standard perturbation theory. In the same section \ref{sec3}, we also discuss issues about the stability of cosmological standard 
perturbation theory and to what extent our matching procedure is independent of those. 
In section~\ref{sec4}, we then explore numerical solutions of viscous fluid dynamics 
based on this matching procedure, and we show that this gives rise to a satisfactory description of large scale structure and its red-shift dependence
in the BAO range up to scale $\sim 0.2\,h/ {\rm  Mpc}$. We finally summarize our main findings in the conclusions. 

\section{Viscous fluid dynamics and cosmological perturbations}
\label{sec2}

The evolution of cosmological perturbations is governed by the Einstein equations
\be
G_{\mu\nu}=8\pi G_N T_{\mu\nu}\, .
\label{einstein} \ee
These equations imply the conservation of the energy-momentum tensor 
\be
\nabla_\nu T^{\mu\nu}=0\, ,
\label{const} \ee
which for the case of an ideal fluid without  any conserved charges  completely defines its dynamics.
 Here, we seek a more realistic (generic) description of the cosmological medium by considering 
viscous-fluid dynamics  as an effective theory of cosmological perturbations
of sufficiently long wavelengths. This approach goes beyond the perfect fluid approximation by accounting 
for deviations from local equilibrium up to first order in gradients, and it allows for non-vanishing
shear and bulk viscosity. Imperfect fluids have been considered in the past for different aspects of cosmology at large scales
(an incomplete list of relevant publications includes \cite{Gagnon:2011id,Sawicki:2012re,Ballesteros:2014sxa,Velten:2013pra,Capela:2014xta,Mirzagholi:2014ifa,Majerotto:2015bra,Maartens:1995wt,Zimdahl:1996ka}).
While the main effort in these works has been the  implications for dark energy or linear behaviour of perturbations, we will be more interested in the growth 
of structure in the mildly non-linear 
evolution of dark-matter perturbations.

The energy-momentum tensor for a relativistic viscous fluid
is of the form 
\be
T^{\mu\nu}=\rho u^\mu u^\nu+(p+\pi_{b}) \Delta^{\mu\nu}+\pi^{\mu\nu}.
\label{tmn} \ee
Here $\rho$ is the energy density and
$p$ the pressure in the fluid rest frame, $\pi_b$ the bulk viscous pressure, and 
$\pi^{\mu\nu}$ the shear-viscous tensor, satisfying: 
$u_{\mu} \pi^{\mu\nu} = \pi^\mu_{~\mu} = 0$. 
The matrix $\Delta^{\mu\nu}$ projects to the subspace orthogonal to the fluid velocity: $\Delta^{\mu\nu} = g^{\mu\nu} + u^{\mu} u^{\nu}$. 
The bulk viscous pressure is
\be
\pi_b=- \zeta  \nabla_{\mu} u^{\mu},
\label{bulkvi} \ee
where $\zeta$ is the bulk viscosity, while the shear viscous tensor is given by 
\begin{equation}
\pi^{\mu\nu} = - 2 \eta \sigma^{\mu\nu} =
-2 \eta\left( \frac{1}{2}\left( \Delta^{\mu\alpha} \nabla_{\alpha} u^{\nu} + \Delta^{\nu\alpha} \nabla_{\alpha} u^{\mu} \right)-\frac{1}{3}\Delta^{\mu\nu}(\nabla_{\alpha}u^{\alpha}) \right),
\label{sigmaa} \end{equation}
with shear viscosity $\eta$. We work in the first-order formalism of relativistic fluid dynamics and drop terms of second order in gradients. In the non-relativistic limit this approximation leads to the standard Navier-Stokes theory. This should be sufficiently accurate for cosmological perturbations well inside the horizon and that relaxation times and other second-order terms would only lead to minor quantitative modifications, although they are in principle needed for a viable causal structure and for linear stability \cite{Israel:1979wp,Hiscock:1985zz}.

The viscous fluid dynamic equations (\ref{const}) take then the explicit form
\bea
u^\mu \nabla_\mu \rho + (\rho+p) \nabla_{\mu} u^{\mu} -
\zeta \left(\nabla_\mu u^\mu \right)^2 -2\eta \sigma^{\mu\nu}\sigma_{\mu\nu}
 & = &0\, ,
\label{hydro1}  \\
(\rho + p+\pi_b) u^\mu \nabla_\mu  u^{\alpha} + \Delta^{\alpha\mu} \nabla_{\mu} (p+\pi_b) 
+ \Delta^{\alpha}_{\;\;\nu} \nabla_{\mu} \pi^{\mu\nu} & = &0\, .
\label{hydro2}
\eea
The Einstein equations (\ref{einstein}) connect perturbations in the matter fields to metric perturbations. The latter enter (\ref{hydro1}), (\ref{hydro2}) via the
covariant derivative $\nabla_\mu$, the projector $\Delta^{\mu\nu} = g^{\mu\nu} + u^{\mu} u^{\nu}$, and the form of the four-velocity $u^\mu=dx^\mu/\sqrt{-ds^2}$.
For the metric, we consider an ansatz of the form
\be
ds^2=a^2(\tau)\left[
-\left(1+2\Psi(\tau,\bx) \right)d\tau^2
+\left(1-2\Phi(\tau,\bx) \right) d\bx\, d\bx \right]\, ,
\label{metric} \ee
that accounts for the dominant scalar metric perturbations only. These are parametrized by the Newtonian potentials
$\Phi$ and $\Psi$. From the linearized  spatial components of the Einstein equations, one can check that the 
difference $\Phi - \Psi$ is proportional to shear viscosity, see eqs. (\ref{pphi}), (\ref{ppsi}) below. 
Keeping two Newtonian potentials is thus necessary for the
description of a fluid with non-vanishing shear viscosity. Since the flow field $u^\mu$ is normalized, $u^\mu u_\mu = -1$, 
it can be expressed through the coordinate velocity $v^i=dx^i/d\tau$ ($i,j=1,2,3$) and the potentials
$\Phi$ and $\Psi$:
\be
u^\mu=\frac{1}{a\sqrt{1+2\Psi-(1-2\Phi)\vec{v}^2}}(1,\vec{v}).
\label{covvel} \ee 
In the next subsection we turn to an analysis of the fluid dynamic equations to linear order in perturbations. Within this framework,
we discuss in subsection~\ref{sec2.2} how the growth of density perturbations is modified by pressure and viscosity. 
In subsection~\ref{sec2.3}, we extend our analysis to the dominant non-linear contributions at the scales of interest.

\subsection{Linear analysis}
\label{sec2.1}

 We parameterize the density and pressure in terms
of spatially homogeneous and isotropic background fields $\rhoz(\tau)$, $\pz(\tau)$ and small spatially varying perturbations,
$\rho(\tau,\bx)=\rhoz(\tau)+\drho(\tau,\bx)$ and $p(\tau,\bx)=\pz(\tau)+\dpp(\tau,\bx)$, with $\drho, \dpp \ll \rhoz$. 
We consider a flat Universe with cosmological constant $\Lambda$ and one species of viscous matter. With the help
of the definitions
\begin{eqnarray}
\Hc&=&\frac{\dot{a}}{a}\, ,
~~~~~~~~~~~~
H=\frac{1}{a} \Hc\, ,
\label{hh} 
\label{hubble} \end{eqnarray}
the equations for the background fields are 
\begin{eqnarray}
&& \Hc^2=\frac{\dot{a}^2}{a^2}=\frac{8\pi}{3}G_N(\Lambda + \rhoz) a^2\, ,
\label{hcdef} \\
&& \dot{\rho}_m+3(\rhoz+\pz)\Hc =0\, .
\label{drhomm}
\end{eqnarray}
We consider perturbations in the variables $\drho$, $\dpp$, $v^i$, $\Phi$, $\Psi$ up to linear order, and we focus on scalar perturbations only.
To expand the fluid dynamic equations in dimensionless measures of these perturbations, we use 
the normalized density contrast $\delta$ and the velocity divergence $\theta$,
\begin{eqnarray}
\delta&\equiv&\frac{\drho}{\rhoz}\, ,
~~~~~~~~~~ 
\theta\equiv\vec{\nabla}\vec{v}\, .
\label{thet} \end{eqnarray}
Also, $\dpp$ is not an independent perturbation, but is related to density perturbations by the equation of state. 
Specializing to a simple equation of state, we write 
\begin{eqnarray}
w&=&\frac{\pz}{\rhoz}\, ,
~~~~~~~~~~
c^2_s = \frac{\dpp}{\drho}\, ,
~~~~~~~~~~
c^2_{ad}=\frac{\dot{p}_m}{\dot{\rho}_m}=w-\frac{\dot{w}}{3(1+w)}\Hc\, ,
\label{cad}  
\end{eqnarray}
where $c_s$ is the velocity of sound. Finally, we introduce shorthands for the normalized so-called kinematic viscosities
\begin{eqnarray}
\label{eq:viscs}
\nu&=&\frac{\eta}{(\rhoz+\pz)a}=\frac{\eta}{(1+w)\rhoz a}\, ,
~~~~~~~~~~
\nu_\zeta = \frac{\zeta}{(1+w)\rhoz a}\, .
\label{deff} \end{eqnarray}

For notational economy, the dependence of the evolution equations on bulk viscosity is not made explicit in our discussion.
However, it can be restored by remembering that bulk viscosity $\zeta$ enters the linearized fluid dynamic equations in 
only two ways. First, it contributes to
the effective pressure of the system, where its contribution can be restored by replacing $p_m \to p_m - 3 \zeta\, H$ in our expressions.\footnote{If one
allows for a density dependence of $\zeta$, this bulk viscous pressure modifies not only the background equation
(\ref{drhomm}), but also the derivatives in (\ref{cad}). We note that, in principle, $p_m - 3 \zeta\, H$ can be negative, see e.\ g.\ ref.~\cite{Gagnon:2011id} and references therein. 
Our formalism can be adapted to this case, but we do not discuss this point further. }
Second, bulk viscosity arises in terms that describe the dissipative attenuation of perturbations. For scalar perturbations it appears 
in the combination $\tfrac{4}{3}\nu + \nu_\zeta$, and can be made explicit by replacing $\tfrac{4}{3} \nu \to \tfrac{4}{3}\nu + \nu_\zeta$. With these provisos, evolution equations for cosmological perturbations in bulk and shear viscous 
matter can be written in Fourier space as
\begin{eqnarray}
\dot{\delta}_{\bk}&=&-(1+w)\left(\theta_{\bk}-3\dot{\Phi}_{\bk} \right)-3\Hc (c^2_s-w) \delta_{\bk}\, ,
\label{dotd} \\
\dot{\theta}_{\bk}&=&-\Hc \theta_{\bk}+k^2\Psi_{\bk}+\frac{c^2_s}{1+w}k^2\delta_{\bk}+3c^2_{ad}\Hc \theta_{\bk}-\frac{4}{3}k^2 \nu \theta_{\bk}\, .
\label{dottheta} \end{eqnarray}
Here, we have introduced the Fourier-transformed fields
\bea 
\delta(\tau,\bx)= \int d^3\bk\,\, e^{i\bk\bx} \td_{\bk}(\tau)\, , \quad \theta(\tau,\bx)&=&\int d^3\bk \,\, e^{i\bk\bx}\theta_\bk(\tau)\, .
\label{fts} \eea
The equations of motion (\ref{dotd}), (\ref{dottheta}) can be closed by the Poisson-like equations for the Newtonian potentials, 
\begin{eqnarray}
k^2 \Phi_{\bk}&=&-\frac{3}{2}\Omega_m\Hc^2\left(3(1+w)\frac{\Hc^2}{k^2}\frac{\theta_{\bk}}{\Hc}+\delta_{\bk} \right)\, ,
\label{pphi}  \\
k^2 \Psi_{\bk}&=& -\frac{3}{2}\Omega_m\Hc^2\left(3(1+w)\frac{\Hc^2}{k^2}\frac{\theta_{\bk}}{\Hc}+\delta_{\bk} +4\nu \theta_{\bk} \right)\, .
\label{ppsi}
\end{eqnarray}

For the case of a pressureless ideal fluid, $w=c^2_s=c^2_{ad}=0$ and $\nu=0$, there is only one Newtonian potential, and the
equations (\ref{dotd}), (\ref{dottheta}) simplify. To understand these simplifications and how they extend to a non-ideal fluid, we discuss now the scales in the problem. The 
 time-scale of the evolution is set by the scale $\Hc$, and we consider subhorizon perturbations with wavenumbers $k\gg \Hc$. 
A time derivative is thus equivalent to a multiplicative factor of $\Hc$, and  it is consistent to assume that the Fourier-transformed fields 
$\delta_{\bk}$ and $\theta_{\bk}/\Hc$ are of comparable magnitude. Eqs. (\ref{pphi}), (\ref{ppsi}) indicate that $\Phi_{\bk}$ and $\Psi_{\bk}$ 
are suppressed by $\Hc^2/k^2$ relative to $\delta_{\bk}$ and $\theta_{\bk}/\Hc$. The above equations then read
\begin{eqnarray}
\dot{\delta}_{\bk}&=&-\theta_{\bk} + {\cal O}\left(\Hc^2/k^2 \right)\, , \quad \qquad \quad \quad\quad\;\;\;\quad \hbox{(ideal fluid)}
\label{dotd1} \\
\dot{\theta}_{\bk}&=&-\Hc \theta_{\bk}+k^2\Psi_{\bk} + {\cal O}\left(\Hc^2/k^2 \right)\, , \quad \qquad \;\;\hbox{(ideal fluid)}
\label{dottheta1}\\
k^2 \Phi_{\bk}&=&k^2 \Psi_{\bk}=-\frac{3}{2}\Omega_m\Hc^2\delta_{\bk} + {\cal O}\left(\Hc^2/k^2 \right)\, . \quad \hbox{(ideal fluid)}
\label{pphi1}
\end{eqnarray}
To study within this linear analysis the late time evolution of the matter distribution, we shall allow
$\delta_{\bk}$ and $\theta_{\bk}/\Hc$ to take values up to order 1.
On the other hand, it is apparent that a perturbative analysis cannot be valid for $\delta_{\bk}, \theta_{\bk}/\Hc \gg 1$. 

The presence of non-zero pressure and viscosity can disrupt the hierarchy of scales that we assumed. In particular, for $c^2_s, w, \nu \Hc$ of order 1,
the terms $\sim k^2 \delta_{\bk}$ and $\sim k^2 \theta_{\bk}$ would dominate in eq. (\ref{dottheta}). They would result in an evolution with a 
characteristic times-scale much shorter than $1/\Hc$, related to the appearance of density waves or exponential damping of fluctuations
through viscosity. In principle, such strong attenuations are conceivable. Here, however, we aim at maintaining an effective description with $1/\Hc$ 
as the only relevant timescale. This corresponds to the {\it assumption}  that 
$w, c^2_s, c^2_{ad}, \nu\Hc \sim \Hc^2/k^2_m$, where $k_m$ is the largest wavenumber (corresponding to the shortest length scale)
for which this effective description is applicable. To make this point explicit, we introduce the parametrization
\begin{eqnarray}
c^2_s&=&\alpha_s \frac{\Hc^2}{k^2_m}\, ,
\label{as} \\
\nu\Hc&=&\frac{3}{4}\alpha_\nu\frac{\Hc^2}{k^2_m}\, ,
\label{anu} 
\end{eqnarray}
and we assume that $\alpha_s$, $\alpha_\nu$ are at most of order 1.
We shall confirm that this assumption is consistent with the values obtained by
 matching viscous fluid dynamics to results from standard perturbation theory, cf. sec.~\ref{sec3}. We shall also provide numerical evidence
that values of this order provide a good description of the power spectrum in the BAO range. 
The linearized equations then become 
\begin{eqnarray}
\dot{\delta}_{\bk}&=&-\theta_{\bk}+ {\cal O}\left(\Hc^2/k^2 \right) \, ,
\label{dotd2} \\
\dot{\theta}_{\bk}&=&-\Hc \theta_{\bk}+k^2\Psi_{\bk}+ \alpha_s \frac{\Hc^2}{k^2_m} k^2\delta_{\bk}
-k^2 \alpha_\nu \frac{\Hc^2}{k^2_m} \frac{\theta_{\bk}}{\Hc}+ {\cal O}\left(\Hc^2/k^2 \right)\, ,
\label{dottheta2}\\
k^2 \Phi_{\bk}&=&k^2 \Psi_{\bk}=-\frac{3}{2}\Omega_m \Hc^2\delta_{\bk}+ {\cal O}\left(\Hc^2/k^2 \right) \, ,
\label{pphi2}
\end{eqnarray}
where we have kept again terms up to ${\cal O}\left(\Hc^2/k^2 \right)$ only. We note that up to this order, effects from finite pressure other than parametrized by the velocity of sound $c_s^2$ do not enter since they are not enhanced by factors $k^2$ in equations (\ref{dotd}) and (\ref{dottheta}). In the following, we sometimes refer to the sound velocity term as effective pressure.

We are now in a position to discuss to what extent effective viscosity and pressure terms can delay or inhibit the growth of structure. To this
end, we combine the linearized equations (\ref{dotd2}), (\ref{dottheta2}) and (\ref{pphi2}) into a single second-order differential 
equation for the density contrast $\delta_\bk$
\be
\ddot{\delta}_\bk+\left(\Hc+\frac{4}{3}k^2 \nu \right) \dot{\delta}_\bk
-\left( \frac{3}{2}\Omega_m\Hc^2-c^2_s k^2    \right)\delta_\bk=0\, .
\label{dddelta} \ee
Changing the evolution variable to $\eta=\ln  a$, this can be written as 
\be
\delta''_{\bk}+\left( 1+\frac{\Hc'}{\Hc}+\alpha_\nu\frac{k^2}{k^2_m} \right) \delta'_{\bk}   -\left( \frac{3}{2}\Omega_m-\alpha_s\frac{k^2}{k^2_m}  \right)\delta_{\bk}    =0\, ,
\label{d2lin} \ee
where the prime denotes a derivative with respect to $\eta$.\footnote{Note that we follow the conventional notation $\eta=\ln(a)$, which should not be
confused with the shear viscosity introduced earlier. We parameterize viscosity by the coefficient $\alpha_\nu$ defined in (\ref{anu}) and (\ref{eq:viscs})
in the following.}

\subsection{The linear growth factor in the presence of viscosity and pressure}
\label{sec2.2}
%
The solution of (\ref{d2lin}) defines the linear growth factor. In the absence of effective pressure
and shear viscosity, this growth factor is independent of the wavelength $2\pi/k$ of the perturbation. For instance, for the simple case of $\Lambda$CDM cosmology, 
and neglecting the radiative components, we have $\Hc'/\Hc=1-3\Omega_m/2$, and, for $\alpha_\nu=\alpha_s=0$, 
the amplitude of the solution takes the well-known form 
\be
D_L(a)=A \frac{\Hc(a)}{a}\int_0^a \frac{da'}{\Hc^3(a')}\, , \quad \hbox{(for $\Lambda$CDM  without radiative component)}
\label{growth} \ee
with $A$ a normalization constant. 

For non-zero effective pressure and shear viscosity, $\alpha_\nu,\alpha_s \not= 0$, eq. (\ref{d2lin}) shows that the growth of cosmological
perturbations depends on their wave vector $k$ \cite{Sawicki:2012re,Velten:2013pra,Capela:2014xta,Majerotto:2015bra}. Moreover, since the effective pressure and viscosity will depend in general on the matter density (and possibly on other characteristics), $\alpha_\nu = \alpha_\nu(\rhoz)$, $\alpha_s = \alpha_s(\rhoz)$, we expect a non-trivial scale dependence
$\alpha_\nu = \alpha_\nu(\eta)$, $\alpha_s = \alpha_s(\eta)$ that reflects the (effective) properties of dark matter. 

The qualitative behavior of the solutions of eq. (\ref{d2lin}) can be understood by the analytical study of simplified set-ups.  To this end, we consider
an Einstein-de Sitter Universe with $\Omega_m=1$, for which  eq.~(\ref{d2lin}) reduces to 
\be
\delta''_{\bk}+A_\nu \delta'_{\bk}-\frac{1}{4}A_s \delta_{\bk}=0,
\label{d2lin1} \ee
with 
\begin{eqnarray} 
A_\nu&=&\frac{1}{2}+\alpha_\nu \frac{k^2}{k^2_m} ,
\label{capan} \\
A_s&=&4\left(\frac{3}{2}-\alpha_s \frac{k^2}{k^2_m} \right).
\label{capas} \end{eqnarray}
For constant values  of $\alpha_\nu$ and $\alpha_s$, we obtain 
a solution of the form
\be
\delta_{\bk}(\eta)=c_1 \exp\left[ \frac{1}{2}\left(-A_\nu+\sqrt{A_\nu^2+A_s} \right)\eta\right] 
+ c_2  \exp\left[ \frac{1}{2}\left(-A_\nu-\sqrt{A_\nu^2+A_s}\right)\eta\right],
\label{solls} \ee
where $c_1$ and $c_2$ are integration constants.
In the long wavelength limit $k/k_m \to 0$, where effective pressure and viscosity
become unimportant, one recovers from this solution the standard growing mode $\sim \exp(\eta)$ and the decaying mode $\sim \exp(-3\eta/2)$. 
Aside from this consistency check, the wavelength dependence of (\ref{solls}) has several interesting features:
\begin{enumerate}
\item \emph{The case of finite effective pressure without effective viscosity ($\alpha_s > 0$, $\alpha_\nu = 0$):}\\
For increasing wavenumber $k$, $A_s$ decreases, and this reduces the exponential growth to $\sim \exp(d \eta)$
with $d < 1$. For $k^2/k^2_m > 3/(2\alpha_\nu)$, $A_s$ becomes negative, and for the slightly higher wavenumber 
$k^2/k^2_m > 25/(16\alpha_\nu)$, the argument in the square root of (\ref{solls}) becomes negative and the solutions become oscillatory. 
This is what one expects from pressure: on scales on which pressure becomes important, it counteracts the growth of the 
density contrast due to gravitational collapse; the system bounces back and starts oscillating. Wavelengths for which this
happens will not form structure, since the real part of both modes 
in (\ref{solls}) is decaying. 
\item
\emph{The case of finite effective viscosity without effective pressure ($\alpha_s =0$, $\alpha_\nu > 0$):}\\
For the mode $\sim \exp(d\, \eta)$ with $d = \frac{1}{2}\left(-A_\nu+\sqrt{A_\nu^2+A_s}\right)$,  the exponential factor $d$
takes the value of the standard growing mode $d=1$ for vanishing viscosity and it approaches zero as $d \sim 3/(1+2\alpha_\nu k^2/k_m^2) > 0$ in the limit of 
very large wavenumber or viscosity. This mode is always growing, $d > 0$.  Thus, in marked contrast to the case of finite pressure, finite viscosity does not inhibit structure formation at any wavelength, but it slows down the formation
of structure on small scales.
\item
\emph{The case $\alpha_s >0$, $\alpha_\nu > 0$:}\\
For perturbations with $k^2/k^2_m > 3/(2\alpha_s)$, $A_s$ becomes negative and both modes of the solution (\ref{solls}) are decaying. Thus, in a viscous medium as well,
pressure inhibits structure formation on small length scales.  However, for perturbations that are sufficiently far in the UV,
the square root $\sqrt{A_\nu^2+A_s}$ is now real, since the factor $\alpha_\nu {k^2}/{k_m^2}$ enters in quadrature: the oscillations that pure pressure
induces are now overdamped by viscous effects. 
\end{enumerate}

\begin{figure}[!h]
\centering
\includegraphics[width=100mm]{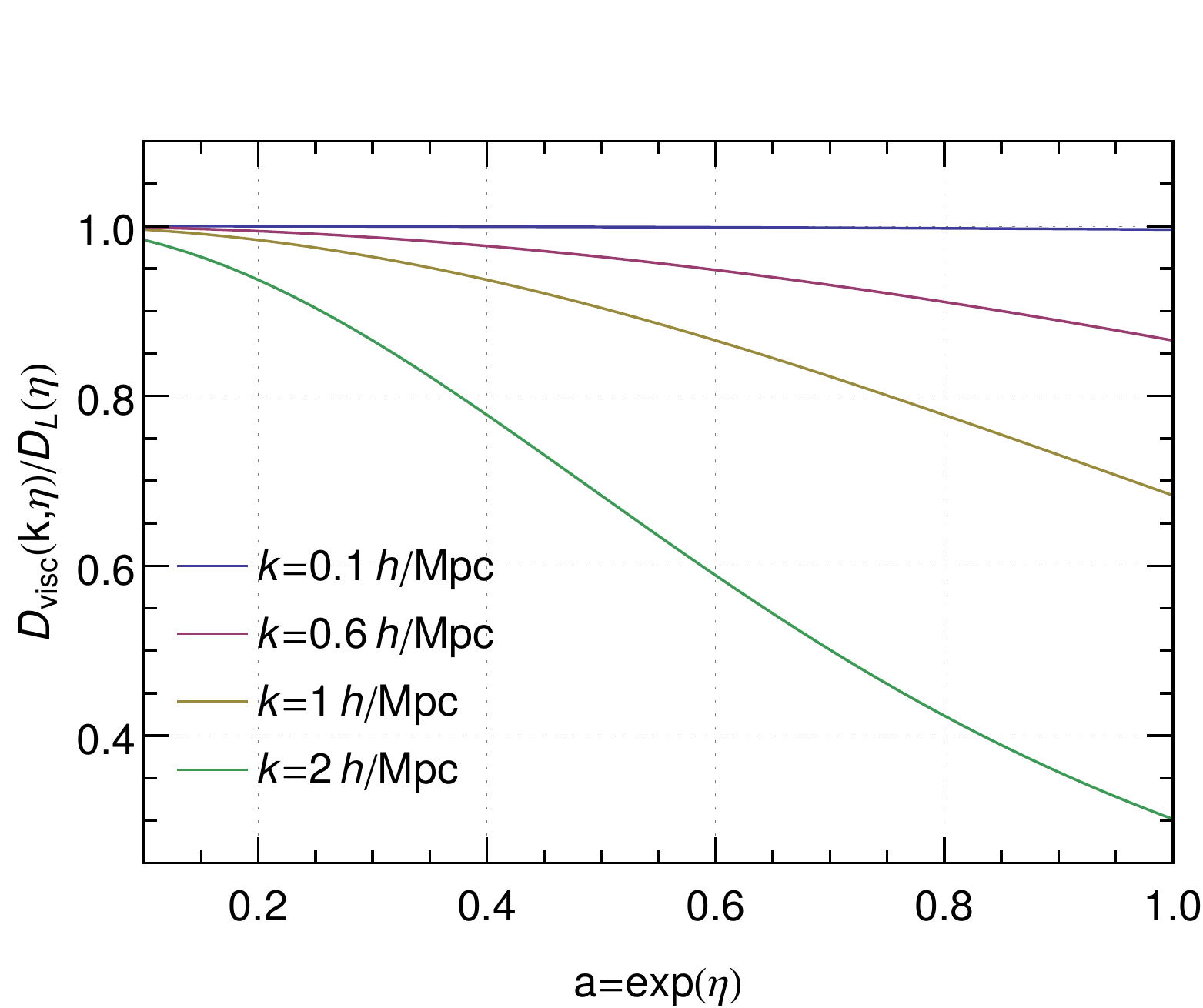}
\caption{The time- and wavenumber-dependence of the linear growth factor $D_\text{visc}(k, z)$ (given by (\ref{sollsHyp}) for $c_1=1$, $c_2=0$) for an Einstein-de Sitter cosmology
in which dark matter is endowed with an effective pressure and viscosity, 
normalized by the corresponding growth factor $D_{L}(z)$ for pressureless non-dissipative dark matter (being identical to the scale-factor $a=e^\eta$ in Einstein-de Sitter).
Shown is the time-dependence for various wavenumbers $k$ assuming  
$\alpha_s, \alpha_\nu \sim  \exp(2 \eta)$, with $k_m=0.6 \, h/$Mpc and $\beta_s = \beta_\nu = 0.67$ (this particular choice will be motivated later in section~\ref{sec3}).
The result also applies to $\Lambda$CDM cosmologies to a very good accuracy when substituting $\eta\to\tilde\eta\equiv\ln( D_L(z))$, and
$\alpha_i\to \tilde\alpha_i$, see eq. (\ref{alphatts}).
}
\label{GrowthD}
\end{figure}

Analytical solutions of (\ref{d2lin1}) can be given even for some time-dependent parameters $\alpha_\nu$, $\alpha_s$. Here, we consider
\be
\alpha_\nu=\beta_\nu \exp(\kappa \eta), ~~~~~~~~~~~\alpha_s=\beta_s \exp(\kappa \eta)\, ,
\label{inter} \ee
for which the solution to (\ref{solls}) can be written in terms of the hypergeometric function $_1F_1(a,b,z)$:
\bea
\delta_{\bk}(\eta)&=&c_1 \exp(\eta)\,
_1 F_1\left(\frac{1}{\kappa}+ \frac{\beta_s}{\kappa\beta_\nu}, 1 + \frac{5}{2\kappa}, -\frac{\beta_\nu}{\kappa} \frac{k^2}{k^2_m}  \exp(\kappa \eta) \right)
\nonumber \\\
&&+c_2 \exp(-3\eta/2)\,
_1 F_1 \left(-\frac{3}{2 \kappa}+ \frac{ \beta_s}{\kappa \beta_\nu}, 1 -\frac{5}{2 \kappa}, -\frac{\beta_\nu}{\kappa}  \frac{k^2}{k^2_m} \exp(\kappa \eta) \right)\, .
\label{sollsHyp} \eea
Of particular interest for the following is the case $\kappa = 2$, since we shall motivate in section~\ref{sec3}
an ansatz with a time-dependence that reduces to eq.\,(\ref{inter}) for this value of $\kappa$. Fig.~\ref{GrowthD} shows the solution
(\ref{sollsHyp}) for $\kappa = 2$ and a set of wavenumbers $k$. One sees that in comparison to the pressureless non-viscous case, 
the linear growth factor is reduced due to finite shear viscosity and pressure. This reduction is very significant for large wavenumbers (small scales), while
for the parameters chosen in fig.\,\ref{GrowthD} scales larger than $\sim 10\, {\rm Mpc}$ are unaffected by the effective pressure and viscosity.  

The main features of the solution (\ref{sollsHyp}) for $\kappa = 2$ are similar to those of the solution (\ref{solls}) for time-independent $\alpha_s$, $\alpha_\nu$. 
In particular, for the case of finite pressure ($\beta_s >0$)
but vanishing viscosity ($\beta_\nu = 0$), the solution of eqs. (\ref{d2lin}), (\ref{inter}) simplifies to
\bea
\delta_{\bk}(\eta)&=&c_1 \exp(-\eta/4)\, J_{-5/4}\left(\sqrt{\beta_s \frac{k^2}{k^2_m} e^{2\eta} } \right) + 
c_1 \exp(-\eta/4)\, J_{5/4}\left(\sqrt{\beta_s \frac{k^2}{k^2_m} e^{2\eta} } \right) .
\label{solls2} \eea
Expanding in the argument of the Bessel functions for small $\beta_s$ (which at fixed time $\eta$ is {\it a fortiori} an expansion for sufficiently long wavelength $2\pi /k$), one 
recovers the standard growing and decaying mode for $\beta_s \to 0$.
On the other hand, it is apparent from (\ref{solls2}) that for sufficiently short wavelength $2\pi/k$, both modes are decaying. Thus, pressure counteracts 
structure formation on small wavelengths, in close similarity to the time-independent case discussed above. 
For the case of finite viscosity ($\beta_\nu > 0$) but vanishing pressure ($\beta_s = 0$), one can expand (\ref{sollsHyp}) for
large viscosity (or equivalently large $k$). For the growing mode one finds in this limit
$\delta_\bk \to c_1 \sqrt{{2}/{\beta_\nu}}\, \left[{\Gamma(\frac94)}/{\Gamma(\frac74)}\right] ({k_m}/{k})$.
Thus the growth is slowed down and asymptotically the linear solution saturates at a finite, scale-dependent value at late times.
This result can be generalized to $\kappa\not=2$,
$\delta_\bk \to c_1 \left({\kappa}/{\beta_\nu}\right)^{1/\kappa} \left[ {\Gamma(1+{5}/{2a})}/{\Gamma(1+{3}/{2a})} \right] \left({k_m}/{k}\right)^{2/\kappa}$.
We conclude that in marked contrast to the physical mechanism induced by pressure, 
finite viscosity does not inhibit structure formation, but it slows it down. 

In the opposite limit of small viscosity/pressure, or equivalently large wave length $2\pi/k$, the leading corrections are for $\kappa=2$ given by
\bea \label{eq:deltaAnalytLongWaveLength}
  \delta_{\bk}(\eta)&\to& c_1 \exp(\eta)\,\left(1-\frac{\alpha_\nu+\alpha_s}{9}\left(\frac{k}{k_m}\right)^2+\frac{3\alpha_\nu^2+4\alpha_\nu\alpha_s+\alpha_s^2}{234}\left(\frac{k}{k_m}\right)^4\right) \\
   && {} + c_2 \exp(-3\eta/2)\,\left(1-\frac{3\alpha_\nu-2\alpha_s}{2}\left(\frac{k}{k_m}\right)^2+\frac{3\alpha_\nu^2+4\alpha_\nu\alpha_s-4\alpha_s^2}{24}\left(\frac{k}{k_m}\right)^4\right)\;. \nn
\eea
As expected, the leading correction scales as $k^2$, and depends only on the sum $\alpha_\nu+\alpha_s$ for the growing mode.
This degeneracy is broken by the $k^4$ corrections, as well as by the decaying mode.

In summary, while the analytical structure of (\ref{sollsHyp}) is more
complex and thus more difficult to discuss than that of the solution (\ref{solls}) for time-independent viscosity and pressure, our short discussion of (\ref{sollsHyp})
points to features of the role of viscosity and pressure that are model-independent.

\subsection{The dominant non-linear terms}
\label{sec2.3}

The complete covariant fluid equations \eqref{hydro1} and \eqref{hydro2}, supplemented by equations for the metric (Einstein's equations) contain many non-linear terms. 
To identify amongst them the dominant ones for the problem at hand, we take recourse to the hierarchy of scales suggested by the linear analysis in section~\ref{sec2.1}. 
More specifically, we make the following assumptions:
\begin{itemize}
\item
We rely on perturbation theory. This means that, during manipulations of the evolution equations,
we assume that the fields $\delta=\drho/\rhoz$, $\theta/\Hc$ and the potentials are small.
However, we allow also for values of order 1 at the limit of validity of our analysis.
\item
With respect to $\Hc/k$,
we treat $\delta$ and $\theta/\Hc$ as quantities of order 1, $\vec{v}$ as a quantity of order $\Hc/k$ and $\Psi,\Phi$ as quantities of
order $\Hc^2/k^2$. 
\item
We assume that a time derivative is equivalent to a factor of $\Hc$, while a spatial derivative to a factor of $k$.
\item
We assume that the couplings
$w, c^2_s, c^2_{ad}, \nu\Hc$ are of order $\Hc^2/k^2_m$, where $k_m$ is the largest wavenumber (corresponding to the shortest length scale)
for which this effective description is applicable. 
\end{itemize}

Within this hierarchy, 
the linear evolution equations given in eqs.~(\ref{dotd2}), (\ref{dottheta2}), (\ref{pphi2}) up to order $\Hc^2/k^2$ receive the following dominant 
non-linear corrections
\bea
\dot{\delta}_{\bk}&+&\theta_{\bk}
+\int d^{3}\bp \,d^{3}\bq\,
\delta^{(3)}(\textbf{k}-\textbf{p}-\textbf{q})\,
 \alpha_1(\textbf{p}, \textbf{q})\, \delta_{\bp}\, \theta_{\bq} 
=0 \, ,
\label{delta} \\
\dot{\theta}_{\bk}&+&\left( \Hc+\frac{4}{3}\nu  k^2 \right)  \, \theta_{\bk}
+\left( \frac{3}{2} \Omega_m \Hc^2-c^2_s k^2 \right) \delta_{\bk}\, ,
\label{theta}  \\
&+&\int d^{3}\bp\, d^{3}\bq\,\delta^{(3)}(\textbf{k}-\bp-\bq)\,
\Bigl({\beta_1}(\bp, \bq)\, \delta_{\bp}\, \delta_{\bq}+ {\beta_2}(\bp, \bq)\, \theta_{\bp}\, \theta_{\bq}
+ {\beta_3}(\bp, \bq)\, \delta_{\bp}\, \theta_{\bq} \Bigr)
=0,
\nonumber
\eea
where~\footnote{Depending on whether we define dimensionless thermodynamic variables by normalizing with respect to $\rhoz$, as done in eqs.
(\ref{thet}), (\ref{cad}) and (\ref{deff}), or with respect to $\rho_m + \delta \rho$, the structure of the couplings (\ref{beta1}), (\ref{beta3}) will change. We shall
demonstrate in section~\ref{sec4} that this ambiguity is numerically unimportant.} 
\bea
\alpha_1(\bp,\bq)&=&\frac{(\bp+\bq)\bq}{q^2}\, ,
\label{alpha1} \\
 \beta_1(\bp,\bq)&=&c^2_s(\bp+\bq)\bq\, ,
\label{beta1} \\
 \beta_2(\bp,\bq)&=&\dfrac{(\bp+\bq)^{2}
\bp\cdot\bq}{2 p^{2} q^{2}}\, ,
\label{beta2} \\
 \beta_3(\bp,\bq)&=&-\frac{4}{3} \nu (\bp+\bq)\bq\, .
\label{beta3} \eea
A particularly useful, compact formulation can be obtained by rewriting these equations for the doublet
\begin{equation}
 \left(
\begin{array}{c}
\phi_{1}(\eta,\textbf{k})\\ \\ \phi_{2}(\eta,\textbf{k})
\end{array}
\right)
=\left(
\begin{array}{c}
\delta_\bk(\tau)\\ \\-\dfrac{\theta_{\bk}( \tau)}{\mathcal{H}}
\end{array}
\right)
\label{doublet}
\end{equation}
and the evolution parameter $\eta=\ln a(\tau)$. One obtains  
\be
\partial_\eta \phi_a (\bk)= -\Omega_{ab}(\bk,\eta) \phi_b (\bk)+ \int d^3 p \, d^3 q \, \delta^{(3)}(\bk-\bp-\bq)
\gamma_{abc}(\bp,\bq,\eta)\, \phi_b (\bp)\, \phi_c(\bq),
\label{eom}
\ee
where
\be
\Omega(\bk,\eta)=\left(
\begin{array}{cc}
~~~0 &~~~-1
\\
-\frac{3}{2} \Omega_m+\alpha_s \frac{k^2}{k_m^2}  &~~~ 1 +\frac{\Hc'}{\Hc}+\alpha_\nu \frac{ k^2}{k^2_m}
\end{array} \right)\, ,
\label{ome} \ee
where the prime denotes a derivative with respect to $\eta$.  The evolution equation (\ref{eom}) is complete up to order $\Hc^2/k^2$
The non-zero elements of $\gamma_{abc}$ are 
\bea
\gamma_{121}(\bq,\bp,\eta)=\gamma_{112}(\bp,\bq,\eta)&=&\frac{\alpha_1(\bp,\bq)}{2}=\frac{(\bp+\bq)\bq}{2q^2} \label{gamma1}, \\
\gamma_{211}(\bp,\bq,\eta)&=&\frac{\beta_1(\bp,\bq)}{\Hc^2}=\alpha_s\frac{(\bp+\bq)\bq}{k_m^2} ,\label{gamma2} \\
\gamma_{222}(\bp,\bq,\eta)&=&\beta_2(\bp,\bq)=\dfrac{(\bp+\bq)^{2} \bp\cdot\bq}{2 p^{2} q^{2}} ,\label{gamma3} \\
\gamma_{212}(\bp,\bq,\eta)=\gamma_{221}(\bq,\bp,\eta)&=&-\frac{\beta_3(\bp,\bq)}{2\Hc}=\alpha_\nu\frac{ (\bp+\bq)\bq}{2k^2_m}.
\label{gamma4} \eea 

In an Einstein-de Sitter Universe, the linear growth factor is $D_L = a$ and the time variable $\eta$ in (\ref{eom}) can thus
be interpreted as $\ln D_L$. In general, this is not the case, but it turns out to be convenient to rewrite the equations of motion
with a modified time variable
\be
\etat=\ln D_L\, ,~~~~~~~~~~~~~~f=\frac{d\etat}{d\eta}= \frac{1}{D_L}\frac{d\,D_L}{d\eta}\, .
\label{ff} \ee
For the suitably redefined doublet 
\begin{equation}
 \left(
\begin{array}{c}
\phit_{1}(\etat,\textbf{k})\\ \\ \phit_{2}(\etat,\textbf{k})
\end{array}
\right)
=\left(
\begin{array}{c}
\delta_\bk(\tau)\\ \\-\dfrac{\theta_{\bk}( \tau)}{\Hc f}
\end{array}
\right)\, ,
\label{doubletr}
\end{equation}
the evolution equations take the form
\be
\partial_\etat \phit_a (\bk)= -\left(   \Omegat^L(\bk,\etat)+\delta\Omegat(\bk,\etat)  \right) \phit_b (\bk)+ \int d^3 p \, d^3 q \, \delta(\bk-\bp-\bq)
\tilde\gamma_{abc}(\bp,\bq,\eta)\, \tilde\phi_b (\bp)\, \tilde\phi_c(\bq),
\label{eomr}
\ee
with 
\be
\Omegat^L(\bk,\etat)=\left(
\begin{array}{cc}
~~~0 &~~~-1
\\
-\frac{3}{2} \frac{\Omega_m}{f^2}&~~~ \frac{3}{2} \frac{\Omega_m}{f^2}-1
\end{array} \right),
~~~~~~~~~~~
\delta\Omegat(\bk,\etat)=\left(
\begin{array}{cc}
~~~0 &~~~0
\\
\alphat_s \frac{k^2}{k_m^2}  &~~~ \alphat_\nu \frac{ k^2}{k^2_m}
\end{array} \right),
\label{omesplit} \ee
and
\be
\partial_\etat=\frac{1}{f}\partial_\eta,~~~~~~~~~~~~~\alphat_s=\frac{\alpha_s}{f^2},~~~~~~~~~~~~~\alphat_\nu=\frac{\alpha_\nu}{f}.
\label{alphatts} \ee
The vertices $\tilde\gamma_{abc}$ are identical to $\gamma_{abc}$ except for the replacement $\alpha_i\to \tilde\alpha_i$.
The advantage of this formulation is that, even for the $\Lambda$CDM cosmology, we have to a very good approximation
\be
\Omegat^L(\bk,\etat)\simeq \left(
\begin{array}{cc}
~~~0 &~~~-1
\\
-\frac{3}{2} &~~~~~~ \frac{1}{2}
\end{array} \right)\, .
\label{omeds} \ee
This is the form of the leading order expression in an Einstein-de Sitter Universe. The transformation to the time variable $\etat$ will thus
allow us in the following to adapt perturbative results derived for an Einstein-de Sitter Universe to a discussion of $\Lambda$CDM cosmology.

\section{Matching effective pressure and viscosity to results from standard perturbation theory}
\label{sec3}

To describe large scale structure with the viscous fluid evolution equations (\ref{eomr}), (\ref{omesplit}), we need to specify the functions
$\alphat_s(\etat)$, $\alphat_\nu(\etat)$ and the scale $k_m$ entering these equations. In the present section, we do this by matching the
fluid dynamic evolution to results for the effective propagator~\cite{CrSc2}
\be
G_{ab}(\bk, \tilde\eta, \tilde\eta^\prime) \delta^{(3)}(\bk - \bk^\prime) = \left\langle  \frac{\delta \phi_a(\bk,\tilde\eta)}{\delta \phi_b(\bk^\prime, \tilde\eta^\prime)} \right\rangle\, ,
\label{eq:defEffProp}
\ee
calculated from standard perturbation theory beyond linear order. This propagator defines the response of the field at late times $\etat$ to a perturbation at 
earlier times $\etat'$. Before presenting technical details of this matching, we discuss first the motivation for this procedure:

As mentioned already in the introduction, viscous fluid dynamics accounts for a wide variety of physically realized systems. 
The viscosities used in these descriptions are not necessarily the {\it fundamental} ones, by which we mean that they do not reflect diffusive transport mediated 
by the fundamental constituents of the liquid. Rather, the dominant source of diffusion is often mediated by fluid dynamic degrees of freedom. 
This is the case in particular if one limits a fluid dynamic description to long wavelengths, while perturbations of short wavelengths
(that are in principle fluid dynamical and whose evolution can be highly non-linear) are not followed explicitly. These UV modes modify the evolution of the IR modes. This modification can be accounted for by introducing scale-dependent effective viscosities and pressure in the fluid dynamic description of
the long-wavelength modes. Examples for this procedure are ubiquitous. They include, for instance, the effective eddy viscosity in cases where laminar flow on large scales 
is modified due to the presence of small-scale turbulent eddies in the system \cite{Frisch}. There are also examples in which the non-linear UV
modes are not vortical degrees of freedom (e.g. wave turbulence). Here, we seek a viscous fluid description of the cosmological fluid on sufficiently 
long wavelengths without following the evolution on short scales. To this end, we aim at choosing $\alphat_s(\etat)$, $\alphat_\nu(\etat)$
as {\it effective} viscosity and pressure in the sense described above. In contrast to fundamental properties of matter, these effective 
properties depend on the scale up to which the effective fluid dynamic description is applied. They also depend on the spectrum of perturbations
above this scale. The matching proposed below will make these features explicit. 

The concept of an effective theory of the low-$k$ modes of the cosmological fluid~\cite{Baumann:2010tm,Carrasco:2012cv,eff1,eff2,eff3,eff4,Manzotti}
is closely related to 
the viscous-fluid dynamic description formulated here. This low-$k$ effective theory contains additional terms
that are not present in  the fundamental description of dark matter. From a Wilsonian point of view, these terms account for the effect of the 
large-$k$ fluctuations that are integrated out in the coarse-grained theory.  In close similarity to the effective material properties in the viscous-fluid dynamical 
description, the couplings in the effective theory depend on the spectrum of UV perturbations and on the cutoff scale 
up to which the effect of these perturbations is coarse-grained. 

In principle, the magnitude and time-dependence of the effective pressure and viscosity can depend on the dynamics of UV modes in a complicated way.
Therefore, in the approach followed in Refs.~\cite{Carrasco:2012cv,eff3,eff4} these quantities were essentially treated as free parameters (or, more precisely, free
functions of time) that have to be determined by a fit to $N$-body data. On the other side,  using $N$-body simulations it was found that the impact of the strongly non-linear physics at small scales on the power spectrum in the BAO range is relatively weak \cite{Bagla:1996zb,nishimichi,Pueblas:2008uv}. 
Motivated by this, we explore in the following the possibility
that for a suitable choice of the coarse graining scale, the dominant contribution to effective viscosity and pressure is generated by modes for
which a perturbative treatment within the fluid picture is still applicable.
This set-up is formulated in the next subsection~\ref{sec3.1},  where we 
match $\alphat_s(\etat)$, $\alphat_\nu(\etat)$ to results from one-loop perturbation theory. In subsection~\ref{sec3.2} we then discuss related issues of
perturbative stability of standard perturbation theory and comment on the possible impact of strongly non-linear dynamics at very short scales.

\subsection{A simple matching procedure to determine effective viscosity and pressure}
\label{sec3.1}

The calculation of the effective propagator (\ref{eq:defEffProp}) in standard perturbation theory provides a particularly simple example of how effects of 
UV modes modify the propagation of long-wavelength cosmological perturbations. Assuming that dark matter is an
ideal pressureless fluid on the fundamental scale, and specializing to the limit of large separation between $\tilde\eta$ and $\tilde\eta^\prime$
(so that only the fastest growing mode needs to be kept), standard perturbation theory yields for small $ k$ and in an Einstein-de Sitter Universe~\cite{CrSc2}
\be
G_{ab}(\bk, \tilde\eta,\tilde\eta^\prime) =g_{ab}(\tilde\eta-\tilde\eta^\prime) +G^{(1)}_{ab}(\bk, \tilde\eta,\tilde\eta^\prime) 
= g_{ab}(\tilde\eta-\tilde\eta^\prime) - k^2 e^{\tilde\eta-\tilde\eta^\prime} \sigma_{d}^2(\tilde\eta) 
\begin{pmatrix} \frac{61}{350} && \frac{61}{525} \\ \frac{27}{50} && \frac{9}{25} \end{pmatrix}\, .
\label{GG}
\ee
Although these results have been derived for the Einstein-de Sitter Universe, they can also be used for $\Lambda$CDM due to the field redefinition (\ref{doubletr}) for the approximation in eq.\ \eqref{omeds}.
Here, the linear propagator is
\be
g_{ab}(\tilde\eta-\tilde\eta^\prime) = \frac{e^{\tilde\eta-\tilde\eta^\prime}}{5} \begin{pmatrix} 3 && 2 \\ 3 && 2 \end{pmatrix} - \frac{e^{-3(\tilde\eta - \tilde\eta^\prime) / 2}}{5} \begin{pmatrix} -2 && ~~2 \\ ~~3 && -3 \end{pmatrix}\, ,
\ee
where the decaying mode in the second term becomes unimportant in the limit of large $\tilde\eta-\tilde\eta^\prime$
for which (\ref{GG}) is written. To lowest order, cosmological perturbations propagate linearly. To one-loop, however, the propagator 
(\ref{GG}) depends on the dimensionful parameter 
\be
\sigma^2_d=\frac{4\pi}{3} \int_0^\infty dq\, P^L(q)\, ,
\label{smallkperp} \ee
where $P^L(k)$ stands for the density-density spectrum. The redshift dependence of $\sigma_{d}^2(\eta)$ in (\ref{GG}) arises from the redshift
dependence of $P^L(q)$. Equations (\ref{GG}), (\ref{smallkperp}) show that in standard perturbation theory at one-loop, a cosmological perturbation 
measured at time $\tilde \eta$ on scale $k$ will be influenced by the spectrum $P^L(q)$ at all scales $q$. 

We seek a viscous fluid dynamic description of the cosmological fluid for scales $k < k_m$ only. Here $k_m$ is a matching scale scale chosen such that
strongly non-linear perturbations have wavenumber larger than $k_m$. Perturbations with $k> k_m$ will not be evolved explicitly, 
but their effect on the long wavelength modes will be parametrized in terms of the effective viscosity and pressure, 
$\alphat_s(\etat)$, $\alphat_\nu(\etat)$. Since the fluid dynamic equations of motion will account explicitly for the interactions  of modes $k< k_m$ with
modes $q < k_m$, we must not absorb such interactions in the definition of $\alphat_s(\etat)$, $\alphat_\nu(\etat)$. These considerations prompt us
to match the expression for the propagator in viscous fluid dynamics to the standard one-loop result (\ref{GG}), with the replacement\footnote{Another formulation that used the propagator to ameliorate the non-linear predictions of 
SPT is renormalized perturbation theory \cite{CrSc1}. This proposal yields a description to resum the infrared modes of the theory and can reproduce the non-linear evolution of the BAO peak \cite{Crocce:2007dt}. Notice that our method makes use of the opposite regime of  UV scales, whose influence we treat in terms of the effective viscosities.}
\be
\sigma^2_d  \longrightarrow \sigma^2_{d, {\rm match}} \equiv  \frac{4\pi}{3} \int_{k_m}^\infty dq\, P^L(q)\, .
\label{smallk} \ee
For simplicity, we use the same matching scale $k_m$ for all redshifts $z$. As a consequence, the $\etat$-dependence of $\sigma_{d, {\rm match}}$ 
results only from the linear growth factor.  This allows us to parametrize $\sigma_{d, {\rm match}}$ as
\be
\sigma_{d, {\rm match}}^2(\etat)=\frac{\beta(k_m)}{2} \frac{D^2_L(\etat)}{k_m^2}\, .
\label{sigmaz} \ee
An estimate for $\beta(k_m)$ can be obtained by computing $\sigma_{d, {\rm match}}$
for a spectrum that scales exactly as $\sim k^{-3}$.
We find
\be
\beta(k_m)= \frac{4\pi}{3}\,k^3_m\,P^L(k_m).
\label{betaapp} \ee
Evaluating $\beta(k_m)$ from this expression results in values near 1 
for $k_m$ in the range $0.4-1\, h$/Mpc.
For our numerical analysis we shall compute $\beta(k_m)$ from the exact linear spectrum.

Our matching procedure implies that we should regard (\ref{GG}) as the solution of an equation of motion for the long-time and long-distance behavior that can be written as

\be
\left( \delta_{ac} \, \partial_\etat + \tilde \Omega_{ac}(\bk,\etat) \right) G_{cb}(\bk, \etat,\etat^\prime) = 0\, .
\label{eq:effectiveLinearEOM}
\ee
We are interested in the limit of this equation for small $\bk$. 
For $\bk \to 0$ the effective propagator approaches the conventional linear propagator and $\Omegat(\bk,\etat) \to \tilde \Omega^L(\tilde \eta)$. 
The leading correction is determined by substituting the one-loop propagator (\ref{GG}) in eq.\ \eqref{eq:effectiveLinearEOM} and solving for
$\Omegat(\bk,\etat)$ to order $k^2$. We find 
\be
\tilde \Omega(\bk,\etat) =\begin{pmatrix} ~~0 && -1 \\ -\frac{3}{2} && ~~\frac{1}{2} \end{pmatrix} 
+ k^2 \sigma_d^2(\etat) \begin{pmatrix} -\frac{3}{175} && - \frac{2}{175} \\ ~~\frac{57}{35} && ~~\frac{38}{35}  \end{pmatrix},
\label{eq:effectiveOmega}
\ee
where we have used that $\sigma_d(\etat)=\exp(\etat-\etat')\,\sigma_d(\etat')$ on the growing mode.
We can now 
compare this expression to the form of the linear $\Omegat(\bk,\etat)$ in the effective theory, 
as given by eq.\ \eqref{omesplit}. 
Apart from numerically small terms in the first row of the second matrix on the right hand side of 
eq.\ \eqref{eq:effectiveOmega} that amount to $\sim 1\%$ corrections, we can match the two expressions if we take 
\bea
\tilde{\alpha}_s &=& \frac{\alpha_s}{f^2}= \frac{57}{35} k_m^2\, \sigma_{d, {\rm match}}^2(\etat)= \frac{57}{70} \, \beta\, D^2_L(z)\, ,
\nonumber\\
\tilde{\alpha}_\nu &=& \frac{\alpha_\nu}{f}= \frac{38}{35} k_m^2\, \sigma_{d, {\rm match}}^2(\etat)= \frac{38}{70}\, \beta \, D^2_L(z)\, .
\label{eq:alphasalphanuMatchedValues}
\eea

 Eqs.\eqref{eq:alphasalphanuMatchedValues} yield separate values for 
$\tilde\alpha_s$ and $\tilde \alpha_\nu$. However, their determination is based on the one-loop expression \eqref{GG} for which only contributions from the growing mode have been kept. Beyond this approximation, the values in  \eqref{eq:effectiveOmega} are expected to change. To make ambiguities in this 
matching apparent, one may therefore project the fluid dynamic expression for $\delta \tilde \Omega(\bk,\tilde \eta) $ on the growing mode,
\be
\delta \tilde \Omega(\bk,\tilde \eta) \cdot \frac{1}{5}\begin{pmatrix} 3 && 2 \\ 3 && 2 \end{pmatrix} =  \frac{1}{5}\begin{pmatrix} 0 && 0 \\ 3(\tilde \alpha_s + \tilde \alpha_\nu) \frac{k^2}{k_m^2} && 2(\tilde \alpha_s + \tilde \alpha_\nu) \frac{k^2}{k_m^2} \end{pmatrix} ,
\ee
and compare it to the second term on the right hand side of \eqref{eq:effectiveOmega}. This shows that the sum $\tilde \alpha_s + \tilde \alpha_\nu$ 
has greater significance than the separate values in \eqref{eq:alphasalphanuMatchedValues}. This sum 
matches exactly the lower row of (\ref{eq:effectiveOmega}). The upper row  of (\ref{eq:effectiveOmega}) cannot be matched in this procedure,
but in comparison to the lower row it is only a 1\% correction. 
\begin{figure}[!t]
\centering
\includegraphics[width=100mm]{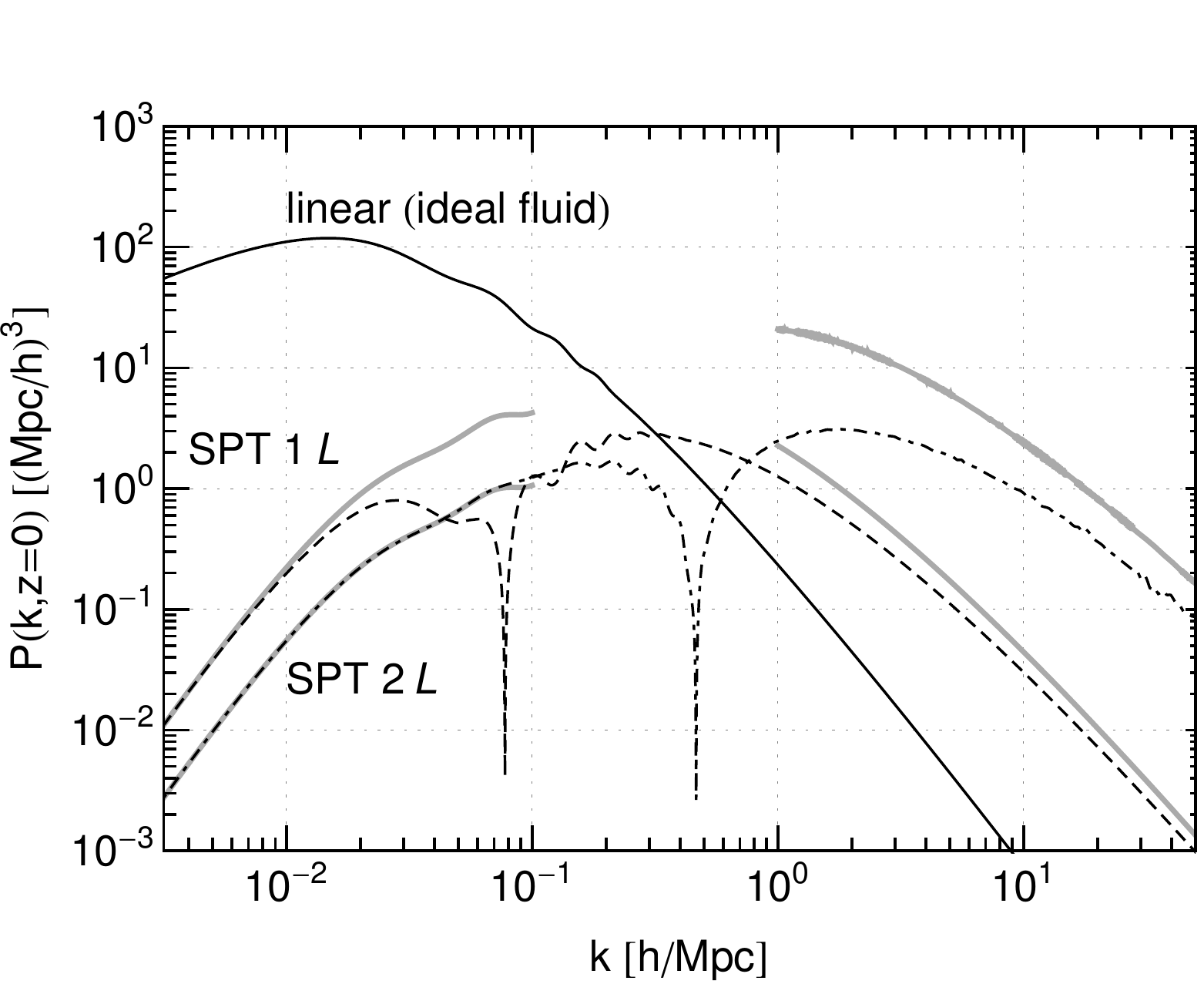}
\caption{Linear matter power spectrum for a $\Lambda${\rm CDM} model with $\sigma_8=0.79$, $\Omega_m=0.26$, $\Omega_b=0.044$, $h=0.72$, $n_s=0.96$,
and the one- and two-loop corrections in standard perturbation theory at $z=0$. The grey shaded lines show analytical approximations valid for
large or small $k$, respectively, see \cite{nonlinear} for details.}
\label{figa}
\end{figure}
%
\subsection{The reliability of the perturbative spectrum and the effective viscosity and pressure}
\label{sec3.2}

In the previous subsection we fixed the effective viscosity and pressure of the fluid dynamic formulation of large scale structure
through eqs.\ (\ref{smallk}), (\ref{eq:alphasalphanuMatchedValues}).  
Here we would like to understand for which range of matching scales $k_m$ one expects this mapping to be accurate,
and furthermore discuss various sources of uncertainty. In particular, (i) the impact of UV modes $k_m< k \lesssim {\cal O}(1\,h/$Mpc$)$
for which the perturbative expansion becomes unreliable, and (ii) the impact of even smaller scales $k\gtrsim {\cal O}(1\, h/$Mpc$)$ for which
the ideal fluid description breaks down.

Our estimate of effective viscosity and pressure terms in sect.\ \ref{sec3.1} is based on one-loop expressions and in particular, $\sigma_{d, {\rm match}}$ contains an integral over the linear power spectrum that extends into the deep UV. It is at this point not clear what corrections $\tilde\alpha_s$ and $\tilde\alpha_\nu$ receive from these modes beyond the one-loop approximation but one might expect terms that resemble the corrections to the power spectrum at low $k$.
The linear density-density spectrum $P^L(q)$ and its one- and two-loop corrections are depicted in Fig. \ref{figa} for $z = \tilde{\eta} = 0$ when non-linear corrections are largest.
It is apparent that the loop corrections are small for low $k$, while they 
become comparable to the linear spectrum in the region $\sim$ 0.4-1 $h$/Mpc, say, and finally dominate for even larger wavenumbers.\footnote{
The spectrum $P^L$ appears in the analytical estimate of the one-loop result for the directly observable matrix of density-density, 
density-velocity and velocity-velocity
correlations~\cite{nonlinear}
\be
P_{ab}^{(1)}(k)=-\left( \begin{array}{cc}
 \frac{61}{105} & \frac{25}{21}
\\
 \frac{25}{21}  & \frac{9}{5}
\end{array} \right)
 k^2 \sigma^2_d P^L(k).
\label{oneloopsmallk} \ee
This expression is valid at all times, indicating that the one-loop correction (and a fortiori our definition of effective viscosity and pressure)
diminishes for growing redshift. A similar expression, involving an additional power of $P^L(k)$, holds for the two-loop correction \cite{CrSc2,nonlinear}. Our discussion in the previous subsection, which was based on the
effective propagator, is equivalent to the analysis of the perturbative spectrum when projected on the growing mode.}
The analytical estimate of the one-loop correction to the 
low-$k$ spectrum, given by eq. (\ref{smallkperp}), involves an
integral that includes the deep UV region. 
Nevertheless, the integral of eq. (\ref{smallkperp}) is dominated by modes $k\ll{ {\cal O}}(1\,h/$Mpc$)$ and the
contribution from the problematic UV region is small. 
The explanation can be traced to the form of the linear
spectrum, which falls off in the UV as $\sim k^{-3}$ (up to logarithmic
corrections), in combination with the decoupling property dictated by total momentum conservation,
which ensures that the impact of UV modes $q$ on the power spectrum at low $k$
is suppressed by a relative factor $(k/q)^2$~\cite{Goroff:1986ep,Rees:1982,Peebles:1980}.
A similar observation can also be made for the
two-loop contribution and in principle for higher loops. Nevertheless, the range of wavenumbers which give the dominant
contribution shifts to higher and higher values with increasing loop order (roughly logarithmically), such that they become
increasingly sensitive to the UV part of the spectrum.
It has been argued in \cite{Blas1} that the perturbative series of
the power spectrum in the low-$k$ limit displays the behavior of an asymptotic series. A resummation
based on Pad\'e approximants suggests that the non-linear corrections to the power spectrum at low $k$ are dominated by the one-loop result,
as one may have naively expected, and correspondingly only mildly dependent on deep UV modes.

It must be emphasized at this point that there is independent evidence from $N$-body simulations that 
the effect of UV modes on the low-$k$ ones through the non-linear mode-mode coupling is weak \cite{Bagla:1996zb,nishimichi}.
This finding is also consistent with an analytical argument that states that small scale structures of size $1/q$ which are virialized 
have a milder impact on larger scales $1/k$ than seen in perturbation theory. In fact, they are suppressed at least as $(k/q)^3$ \cite{Peebles:1980} (see  \cite{blas}
for a recent discussion). Moreover, it was found in \cite{Carrasco:2012cv} that the dependence of the power spectrum extracted from coarse-grained $N$-body
data is consistent with the dependence on $k_m$ obtained from eq. (\ref{smallk}) for low $k_m$. 

The impact of very short scale perturbations, for which the ideal fluid picture breaks down, on the power spectrum at low $k$
has been investigated in \cite{Pueblas:2008uv}. This was done by extracting the second moment of the velocity distribution from $N$-body data. The divergence of this
velocity dispersion contributes on the right-hand side of the Euler equation, and could be interpreted in terms of
effective pressure and viscosity as well. Ref. \cite{Pueblas:2008uv} finds that the impact on the power spectrum is small, below $\%$ level
for $k\lesssim 0.2\, h/$Mpc at $z=0$, which would lead correspondingly to small values $\alpha_s, \alpha_\nu\ll 1$ when interpreted as
effective pressure and viscosity (assuming $k_m$ well below $1 \,h/$Mpc). See also \cite{Valageas:2010rx, Manzotti,Baldauf:2015tla} for related discussions.

Altogether these findings
suggest that, for a coarse-graining scale $k_m$ below ${ {\cal O}}(1\,h/$Mpc$)$, the estimates in the previous section capture the dominant
contribution to the effective pressure and viscosity, with the largest contribution coming from scales close to $k_m$. Indeed,
the integral in eq. (\ref{smallk}), on which the estimate is based, is dominated by the region
near the lower limit $k_m$. It is apparent then
that $k_m$ cannot be taken arbitrarily large, as this would move us into a region in which the
perturbative expansion fails, making the deduced values of viscosity and sound velocity unreliable.
At the practical level, we expect that the matching scale $k_m$ must be taken in the region $0.4-0.8 \,h$/Mpc,
with the upper value lying at the edge of applicability, and the lower
value chosen in a way such that the BAO range is well covered. 
A crucial test of the validity of our assumption is the independence of our final results on
the specific value of $k_m$ within the range discussed above. 
We shall verify this independence in the next
section. 

We point out that our construction does not constitute a mere rewriting 
of standard perturbation theory. The crucial difference lies exactly in the way we deal with
the UV modes. We isolate the effect on the low-$k$ modes arising from modes near and slightly above
$1\,h$/Mpc, and translate it into effective low-energy parameters (viscosity and sound velocity).
We expect that these give the dominant effect, while the influence of the deep UV is negligible.
Nevertheless, it is important to keep in mind that eventually one expects small corrections to the effective pressure and
viscosity. We then use perturbation theory within the low-$k$ theory, cutting off all integrations with
an upper limit equal to $k_m$ and fully taking 
into account the wavenumber and time dependence of the propagator that arises from the effective pressure and viscosity. We expect a quick convergence of the perturbative series, a property
that we shall verify in the following section. 
On the other hand, standard perturbation theory becomes increasingly sensitive to the deep UV at
higher loops, a feature that prohibits its convergence.


\section{Results}
\label{sec4}

Computing non-linear corrections to the power spectrum within viscous fluid dynamics is complicated by
the wavenumber- and time-dependence of the additional pressure and viscosity terms (\ref{omesplit}) in the linear propagation, as well as
additional vertices (\ref{beta1}) and (\ref{beta3}). We use two different methods, which yield results that are in agreement.
One of the methods turns out to be efficient enough for computing the power spectrum up to two loops within the viscous theory,
with time- and wavenumber-dependent propagation.
Before discussing our main numerical results, we briefly describe the methods to compute power spectra
within the viscous theory, as well as the sensitivity to various assumptions.
Our numerical results are based on a $\Lambda${\rm CDM} model with parameters
$\sigma_8=0.79$, $\Omega_m=0.26$, $\Omega_b=0.044$, $h=0.72$, $n_s=0.96$.

\subsection{Numerical solution}
\label{sec4.1}

The first method we use is based on the time-flow approach developed in \cite{time}. It 
is based on coupled evolution equations for the power spectrum and the bispectrum.
They follow in a straightforward way by taking time-derivatives of the power- and bispectra
of density and velocity perturbations, and using the equation of motion (\ref{eomr}).
The non-linear term leads to a coupled hierarchy of evolution equations. The time-derivative of the
power spectrum depends on the bispectrum, and the time-evolution
of the bispectrum depends on the trispectrum, i.e. the four-point function. Within the time-flow approach, the system
of equations is closed by neglecting the connected part of the four-point function.
The bispectrum is then sourced by the disconnected parts, which are
quadratic in the power spectrum.

We adapt the method
described in \cite{time} in several ways: first, we take the wavelength-dependence in the linear propagation
into account in the evolution of the bispectrum, as described in \cite{blas}. Second, we include the additional
non-linear vertices (\ref{beta1}) and (\ref{beta3}). Third, we implement a non-zero initial condition for the bispectrum as described
in \cite{Audren:2011ne, blas}. This is important in order to obtain reliable results when initializing the evolution at a finite redshift,
as is mandatory in practice. Fourth, we solve the time-flow equations using the linear solution for the power spectrum in order to compute the source term in the
evolution equation for the bispectrum. This is equivalent to a one-loop computation, which however takes the time- and
wavenumber-dependence of the propagation into account. As was shown in \cite{Audren:2011ne}, this approximation differs only marginally
from a self-consistent solution of the flow equations for $\Lambda$CDM cosmologies.

To go beyond one-loop accuracy within the time-flow approach, it is necessary to go one order higher
in the hierarchy, i.e. to include the (connected) trispectrum. Indeed, it was recently shown that including the trispectrum is
important to ensure that the resulting higher-order corrections to the bispectrum respect the so-called consistency relations,
which are non-perturbative relations associated with Galilean invariance and with the response to local curvature \cite{Ben-Dayan:2014hsa}.
Unfortunately, solving the time-flow equations with a non-zero trispectrum becomes rather cumbersome, even for
Einstein-de Sitter and a pressureless, ideal fluid description for which significant simplifications are available due to
the wavelength-independent propagation \cite{Juergens:2012ap}.

Since we are interested in a situation with wavelength-dependent propagation, we opt for another approach,
which turns out to be more efficient in this case. This allows us to go to the two-loop level,
taking the wavenumber- and time-dependence in (\ref{omesplit}) into account.
This second method is a more direct generalization of the standard perturbative approach. We expand the
density contrast $\delta\equiv \tilde\phi_1$ and rescaled velocity divergence $-\theta/({\cal H}f)\equiv \tilde\phi_2$
in powers of the initial perturbations at $\tilde\eta=\etat_0$,
\be
  \tilde\phi_a(\bk,\tilde\eta) = \sum_n \int d^3q_1 \cdots d^3q_n \, (2\pi)^3 \delta^{(3)}(\bk-\sum_i\bq_i) F_{n,a}(\bq_1,\dots ,\bq_n,\tilde\eta)
  \delta_{\bq_1}(\eta_0)\cdots\delta_{\bq_n}(\etat_0)\;.
\ee
Inserting this series into the equation of motion (\ref{eomr}) gives an evolution equation for the kernels $F_{n,a}$,
\bea\label{kernels}
  (\partial_{\tilde\eta}\delta_{ab}+\tilde\Omega_{ab}(\bk,\tilde\eta))F_{n,b}(\bq_1,\dots ,\bq_n,\tilde\eta)
  &=& \sum_{m=1}^n \tilde\gamma_{abc}(\bq_1+\cdots+\bq_m,\bq_{m+1}+\cdots+\bq_n) \\
  && \times F_{m,b}(\bq_1,\dots,\bq_m,\tilde\eta)F_{n-m,c}(\bq_{m+1},\dots,\bq_n,\tilde\eta) \;, \nn
\eea
where the right-hand side is understood to be symmetrized w.r.t. arbitrary permutations of the $\bq_i$, and $\bk=\sum_i \bq_i$.
When neglecting the pressure and viscosity terms, the solution coincides with the standard Einstein-de Sitter kernels \cite{bernardeau0},
$F_{n,1}|_{\alpha_\nu=\alpha_s=0}=e^{n\tilde\eta}F_n$ and $F_{n,2}|_{\alpha_\nu=\alpha_s=0}=e^{n\tilde\eta}G_n$. For non-zero
pressure and viscosity, the time-dependence does not factorize and it is hard to do the time-integration analytically. We
therefore solve the differential equations numerically.
Since the effective
pressure and viscosity become negligibly small at early times, we initialize the kernels with their Einstein-de Sitter values at an
early initial time chosen to be $\tilde\eta=-4$ (we checked that our results are stable when varying the starting
time). The equation (\ref{kernels}) allows for a recursive determination,
starting from the linear solution ($n=1$). This solution is then used to determine the kernels with $n=2$, etc.
To obtain the power spectrum at one-loop, one needs to go up to $n=3$, and at two-loop up to $n=5$.

The power spectrum is then obtained by the same expressions as in standard perturbation theory (i.e. $P_{1-loop}=P_{22}+2P_{13}$,
$P_{2-loop}=P_{33}+2P_{15}+2P_{24}$) \cite{bernardeau0, nonlinear}, except that the
kernels are given by the time-dependent solutions of (\ref{kernels}) instead of the Einstein-de Sitter kernels. 
For an efficient implementation it is essential to
compute the kernels only for the wavevector configurations that are actually required for the loop integrals,
e.g. $F_{5,a}(\bk,\bq,-\bq,\bp,-\bp, \tilde \eta)$ for the two-loop contribution $P_{15}$, along with the
lower-order kernels required for the recursive solution. 
Using the manipulations of the integrand and the algorithm described in \cite{Blas1}, this method allows us
to compute the power spectrum up to two loops within the viscous theory at an affordable numerical cost (of order a couple of minutes
per external wavenumber $k$ on a single desktop machine). We checked that the one-loop results obtained with both methods are in excellent agreement.

\begin{figure}[htb]
\centering
\includegraphics[width=100mm]{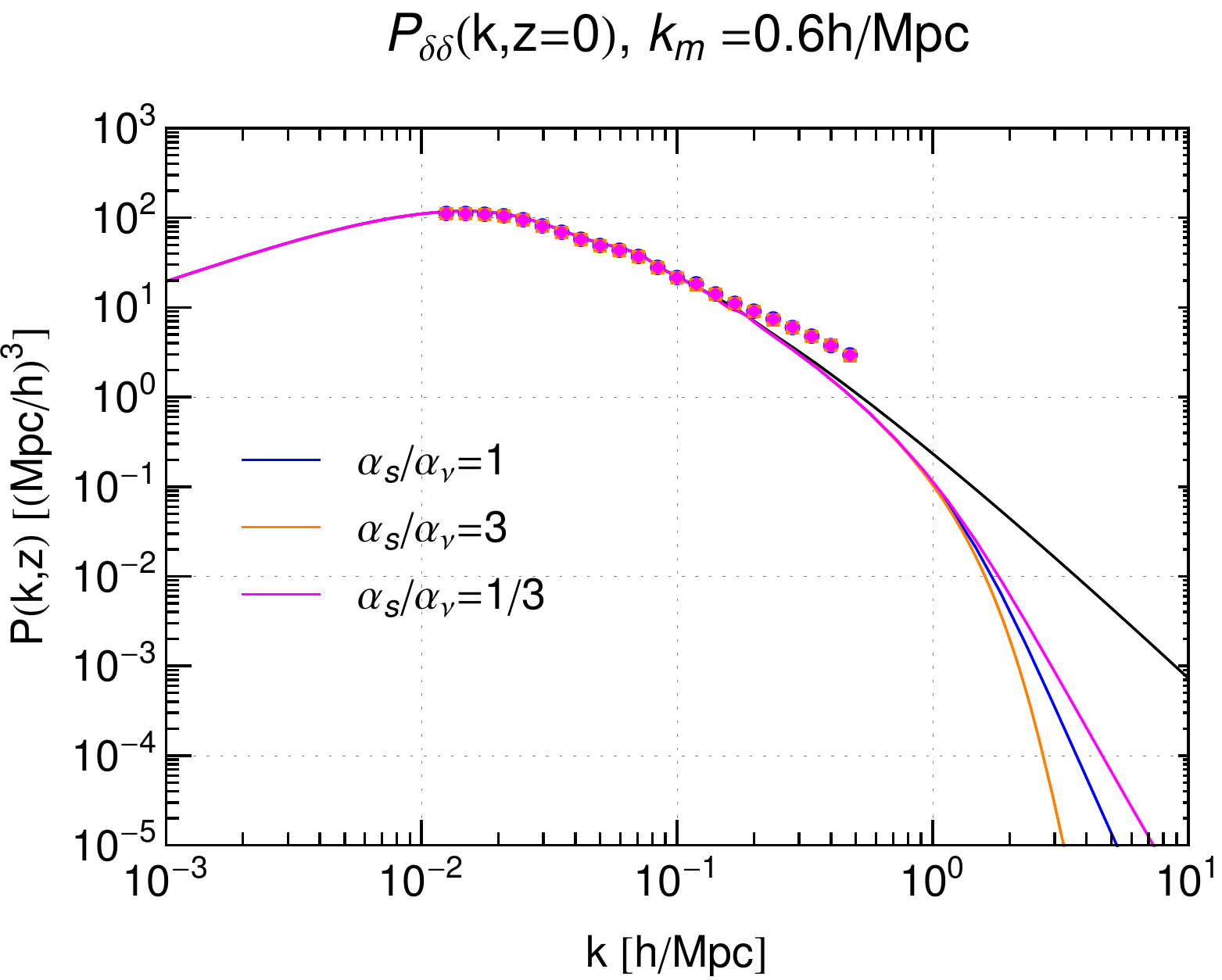}
\caption{
{Power spectra in linear and one-loop approximation for a viscous fluid. The lines (dots) show the linear (one-loop) solution obtained for various ratios
of viscosity and pressure terms. Their sum is kept constant. The differences between the various results are below the $\%$ level for $k<0.6\, h/$Mpc.
For comparison, the black solid line shows the linear spectrum obtained in the pressureless, ideal fluid case.}}
\label{figr}
\end{figure}

\begin{figure}[htb]
\centering
\includegraphics[width=100mm]{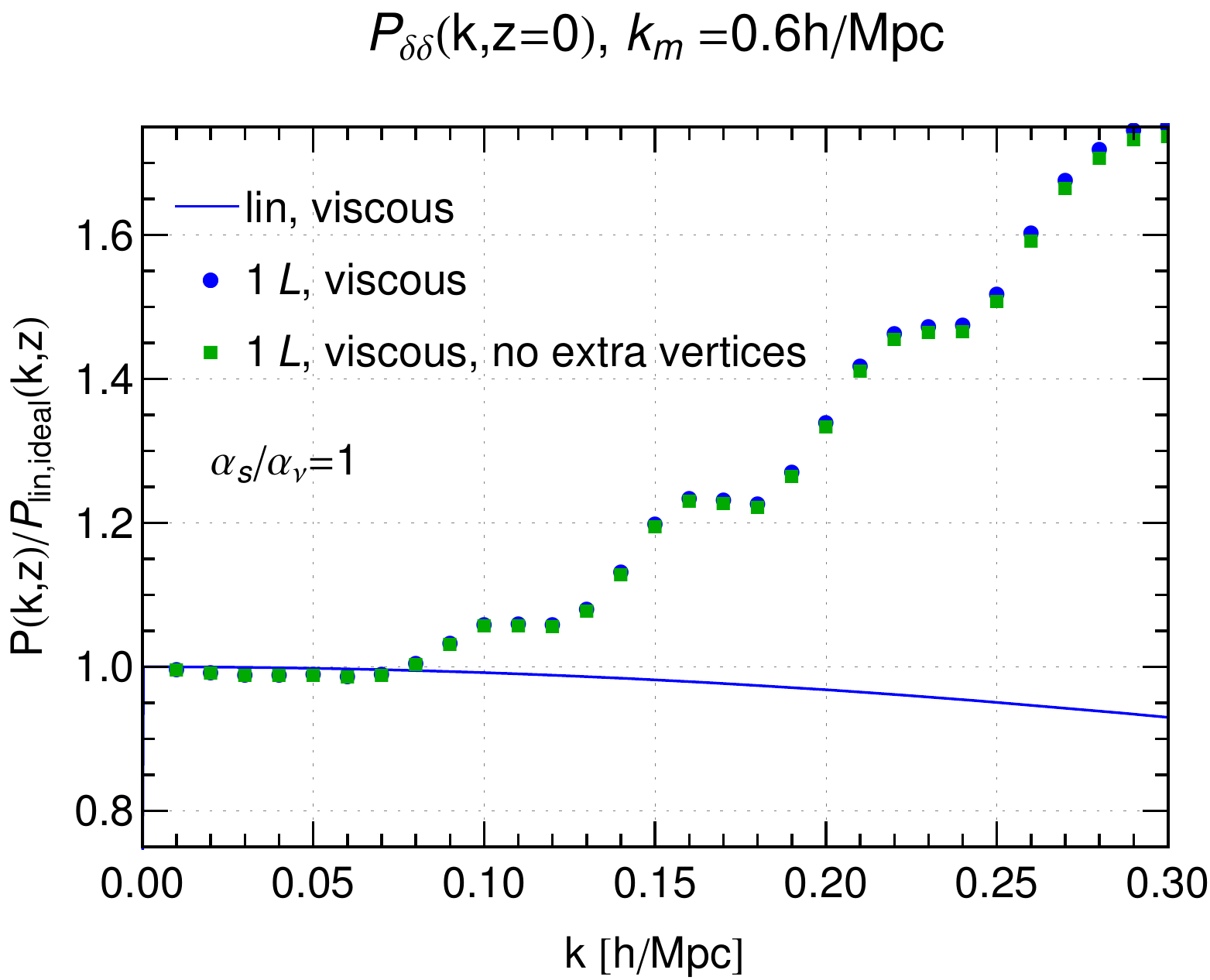}
\caption{
{One-loop power spectrum for a viscous fluid at $z=0$ for two different approximations of the
non-linear coupling terms. The blue dots correspond to the result when taking the standard vertices (\ref{alpha1}) and (\ref{beta2}) and
the extra vertices (\ref{beta1}) and (\ref{beta3}) into account. The green squares show the result when neglecting the extra vertices.}}
\label{figv}
\end{figure}

\subsection{Pressure versus viscosity and extra vertices}

The matching procedure described in subsection\,\ref{sec3.1} allows us to obtain separate expressions for the parameters
$\tilde\alpha_\nu$ and $\tilde\alpha_s$ characterizing the effective viscosity and pressure, respectively.
However, only their sum enters the growing mode of the linear solution in the low-$\bk$ limit. Nevertheless,
the non-linear contributions are sensitive also to decaying modes inside the loops, and to wavenumbers
for which the low-$\bk$ limit is not applicable. Therefore, one may wonder how sensitive the power spectrum
is to changing the ratio $\tilde\alpha_\nu/\tilde\alpha_s$ while keeping their sum fixed according to (\ref{eq:alphasalphanuMatchedValues}).
In fig.\,\ref{figr} we show the dependence on this ratio for a matching scale $k_m=0.6\,h/$Mpc at $z=0$.
While the linear solutions differ for $k\gtrsim 1\,h/$Mpc, we find that both the linear and the one-loop
result for the power spectrum agree to better than 1$\%$ for $k<k_m$. Thus, while pressure and viscosity
lead to qualitatively different behaviour at large $k$, their effect is degenerate within the weakly
non-linear regime. This is consistent with the analytical expression for the linear propagator in eq.\ \eqref{eq:deltaAnalytLongWaveLength} and in a similar form, this feature has also been observed within a perturbative treatment of effective pressure and viscosity terms in \cite{eff3}.

Apart from a more complicated propagator, viscosity and pressure in general also induce additional non-linear
interaction vertices (\ref{beta1}) and (\ref{beta3}). In fig.\,\ref{figv} we show the one-loop solution obtained when taking these vertices
into account, compared with the case where only the standard vertices are kept, again for $k_m=0.6\, h/$Mpc and
at $z=0$. We find that the difference is below the 1$\%$-level for $k<k_m$. 
We therefore neglect these extra vertices in the following. This finding also justifies why it is not
crucial to resolve possible
ambiguities in the precise magnitude of these extra vertices, as discussed in subsection\,\ref{sec2.3}.

\begin{figure}[htb]
\centering
$$
\includegraphics[width=0.5\textwidth]{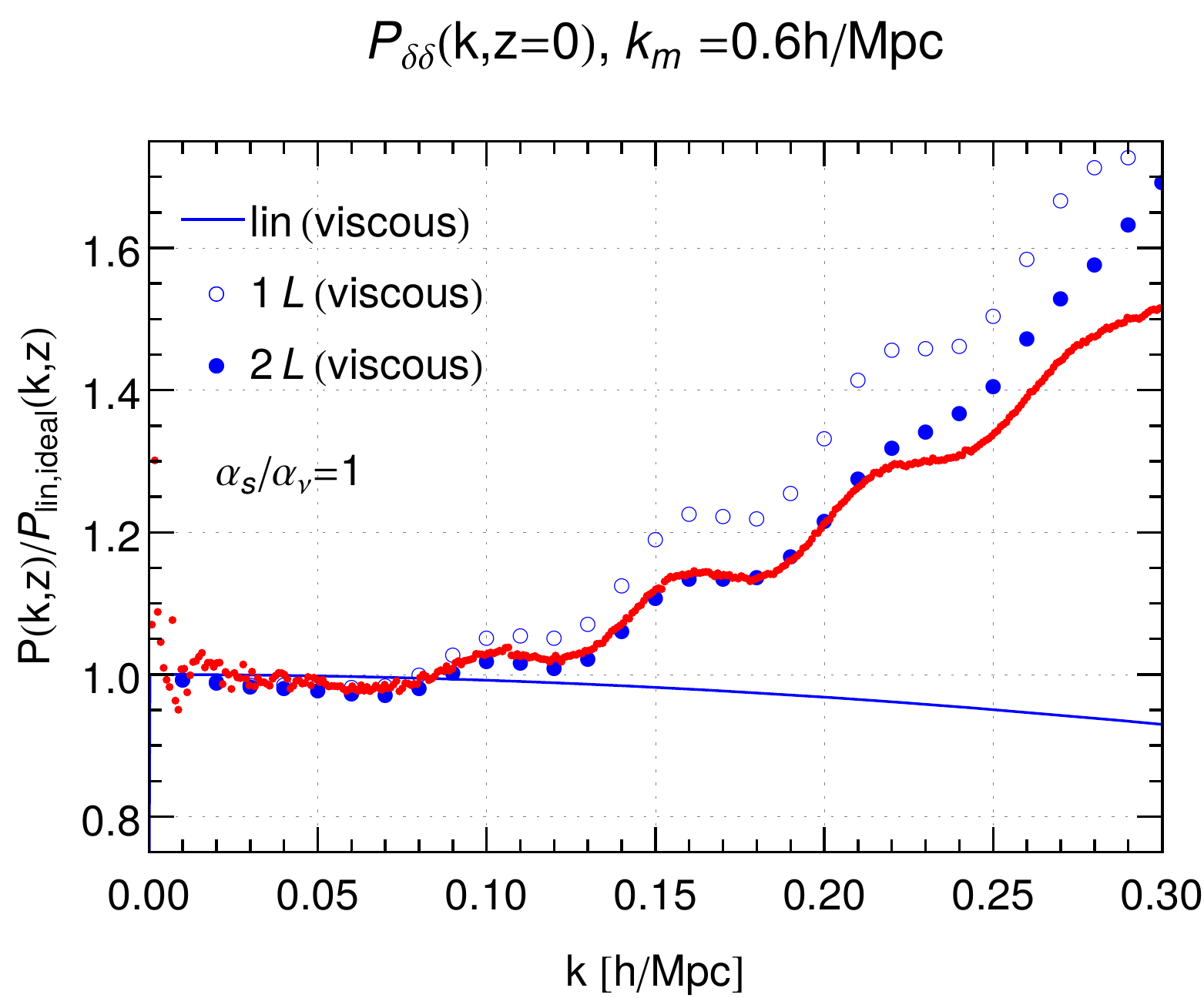}\qquad 
\includegraphics[width=0.5\textwidth]{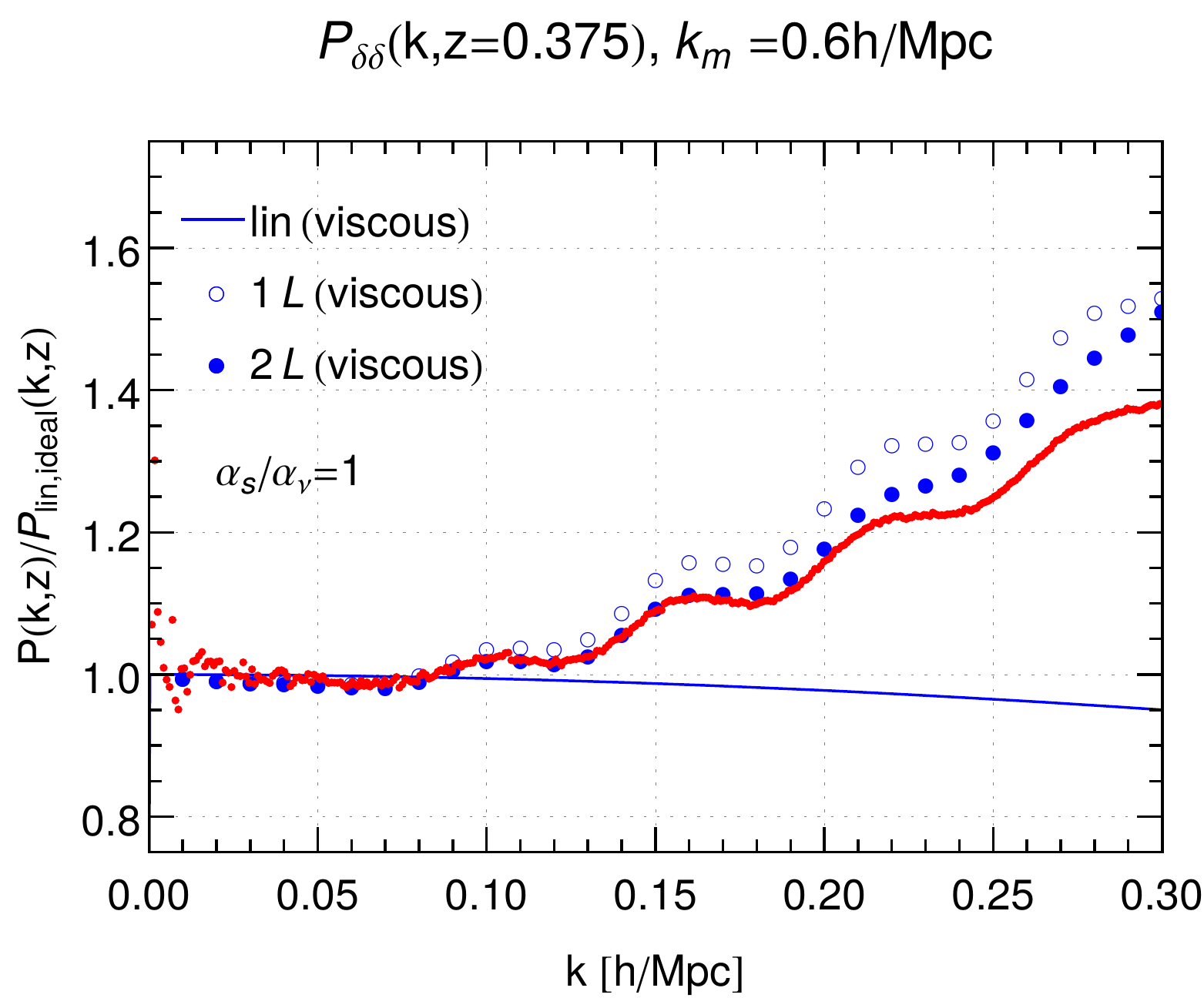}
$$
\\
$$
\includegraphics[width=0.5\textwidth]{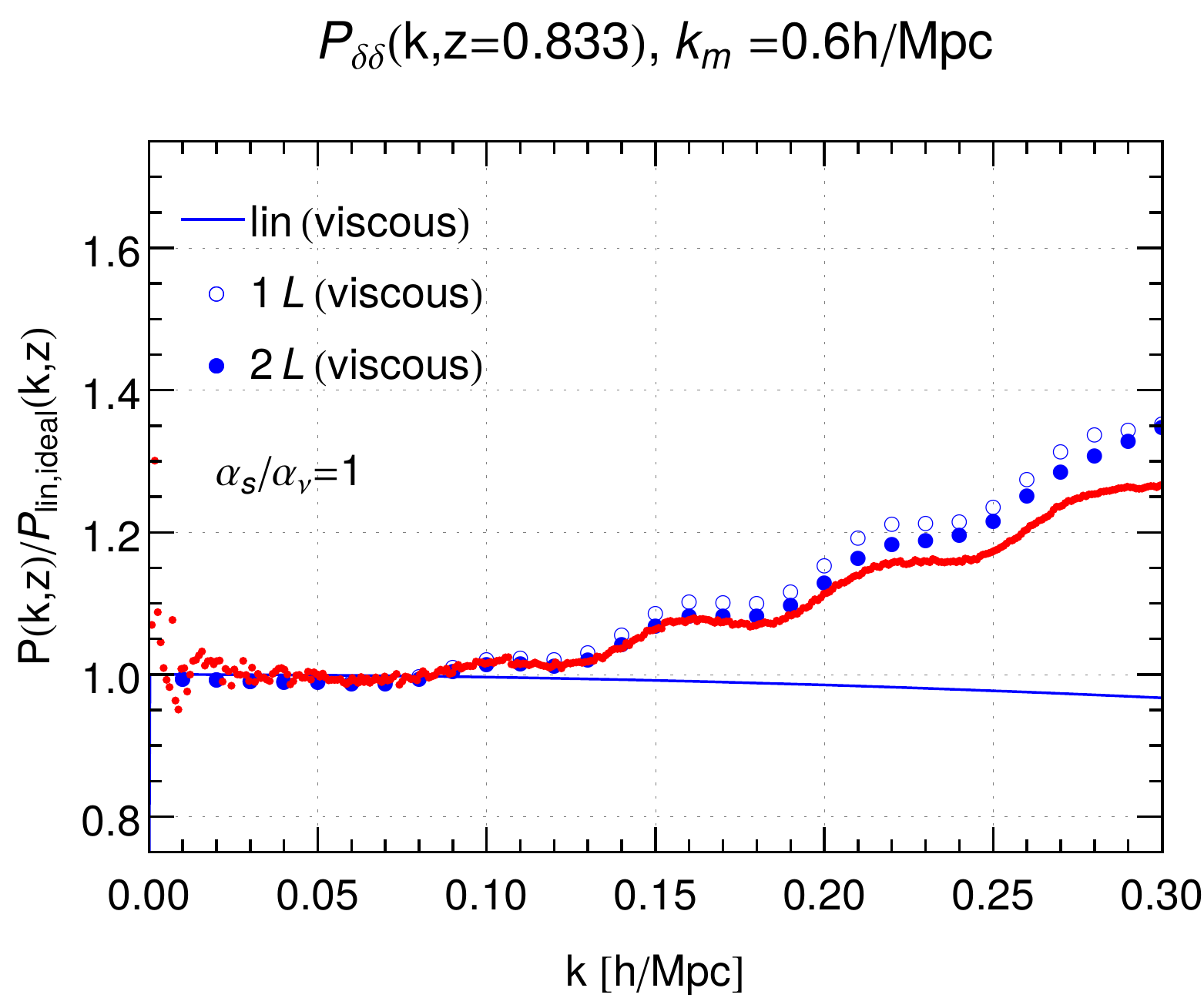}\qquad 
\includegraphics[width=0.5\textwidth]{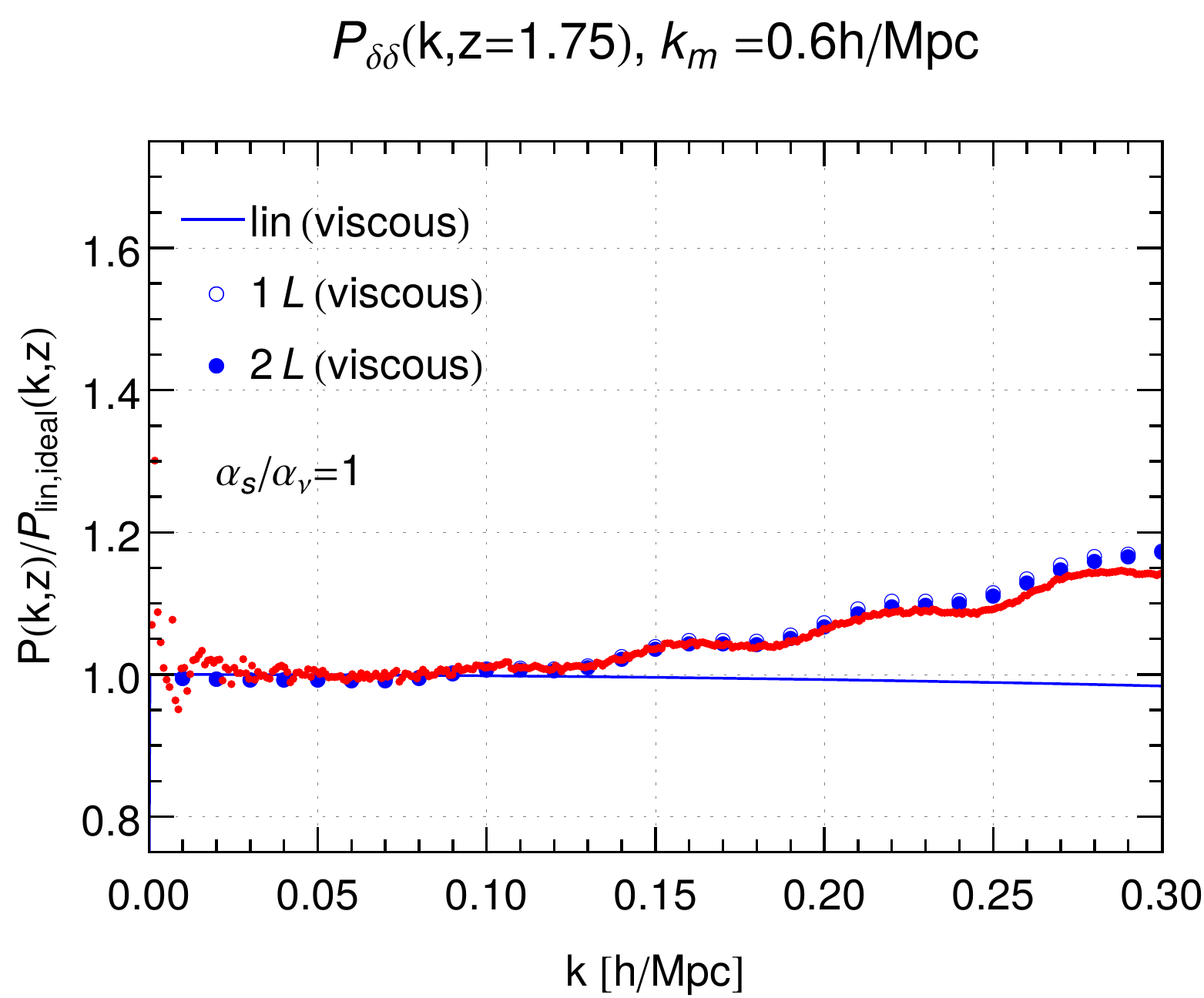}
$$
\caption{
{Power spectrum obtained in the viscous theory for redshifts $z=0$, $z=0.375$, $z=0.833$ and $z=1.75$, respectively, normalized to the conventional
 linear spectrum $P_{lin, ideal}$ (i.e. assuming an ideal fluid). The 
 open (filled) circles show the one-loop (two-loop) result computed in the viscous theory. The solid blue line is the linear spectrum computed in the viscous theory, and the
 red points show results of an $N$-body simulation~\cite{Kim:2011ab}.}}
\label{figk1}
\end{figure}

\subsection{Comparison with $N$-body simulation and dependence on the matching scale}

In fig.\,\ref{figk1} we show the matter power spectrum at various redshifts, computed for a viscous fluid.
All curves are normalized to the linear spectrum obtained for an ideal, pressureless fluid. The solid lines show the
linear spectrum in the viscous theory, which is suppressed compared to the ideal fluid for large $k$ and small $z$,
as expected. The open (filled) circles show the one-loop (two-loop) result, taking the
time- and wavelength-dependent viscosity and pressure terms into account, and including modes with $q<k_m$ in the
loop integrals. The viscosity and pressure are determined from the matching
procedure described in subsection\,\ref{sec3.1} with $k_m=0.6 \, h/$Mpc. 

\begin{figure}[htb]
\centering
$$
\includegraphics[width=0.5\textwidth]{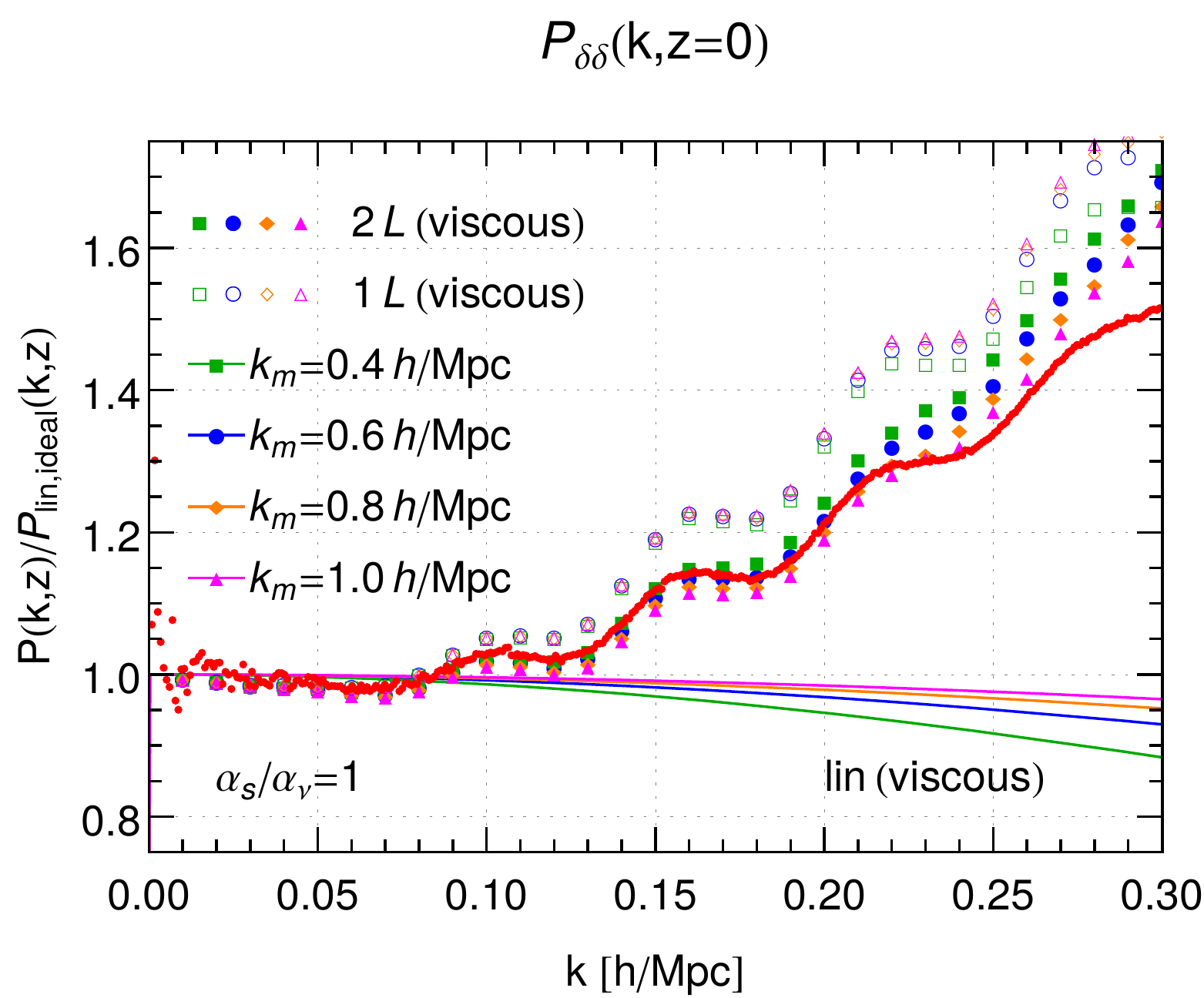}\qquad 
\includegraphics[width=0.5\textwidth]{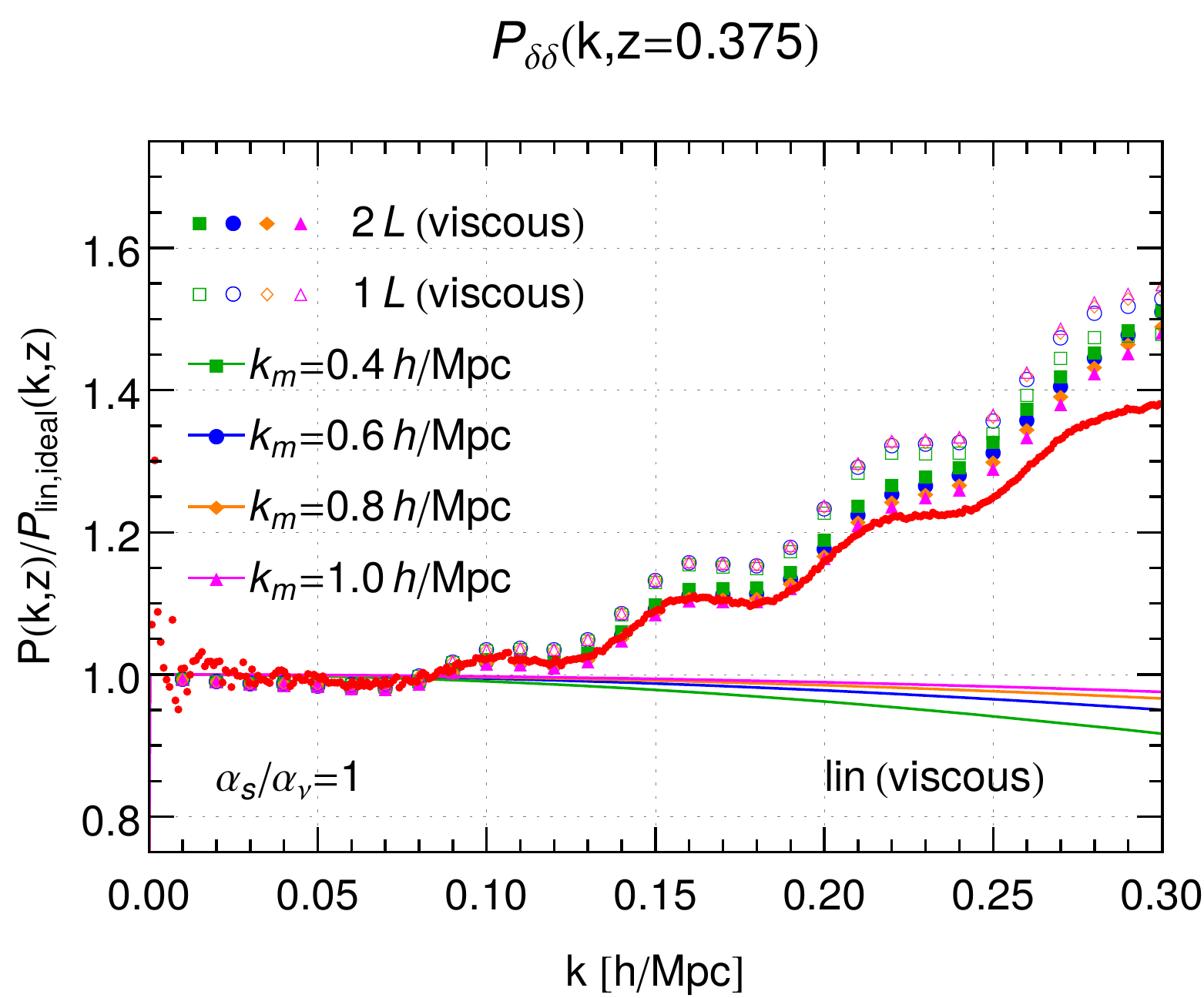}
$$
\caption{
Dependence of the power spectrum on the matching scale $k_m$, for $z=0$ and $z=0.375$. The open (filled) symbols show the one- (two-) loop
results as in fig.\,\ref{figk1}, but for various values of $k_m$. The lines show the linear results. The numerical values of
the effective pressure and viscosity coefficients obtained from the matching condition (\ref{eq:alphasalphanuMatchedValues})
are given by $\tilde\alpha_\nu+\tilde\alpha_s=(1.02, 1.34, 1.60, 1.80)\cdot D_L(z)^2$
for $k_m=(0.4, 0.6, 0.8, 1.0)\, h/$Mpc, respectively.
}
\label{figkm}
\end{figure}

For comparison, the red points show corresponding results from a large-scale
$N$-body simulation \cite{Kim:2011ab}. The two-loop results agree reasonably well with the $N$-body data for $k\lesssim 0.2-0.25 \, h/$Mpc
at low redshift, and the agreement improves at higher redshift. It must be emphasized that no parameter adjustment
has been made in order to reproduce the $N$-body results. 

Since the full result should be independent of the choice of $k_m$, any residual dependence can be taken as a
measure of the theoretical uncertainty. In fig.\,\ref{figkm} we show the dependence on $k_m$ in the range
$0.4-1\, h/$Mpc. The variation of the two-loop results is small within the range for which the results agree well
with $N$-body data. The size of the two-loop correction increases with $k_m$, i.e. also the perturbative
uncertainty becomes larger. For the largest value $k_m=1 \,h/$Mpc, the deviations start to become sizeable,
as expected from the previous discussion. 
As mentioned above, the residual dependence on the matching scale can be taken as an estimate for the theoretical uncertainty. Considering a
variation over a factor of about two, in analogy
to common practice within quantum field theoretical computations, indicates that our results are robust at the $\%$ level for $k<0.2\, h/$Mpc at $z=0$. As discussed in
section~\ref{sec3.2}, the remaining dependence can be attributed to uncertainties in the determination of the effective pressure and viscosity. In addition,
deviations with respect to $N$-body simultations are expected to occur for $k_m\gtrsim 1 \,h/$Mpc or $k\gtrsim 0.2\, h/$Mpc because either the
coarse-graining starts to become
sensitive to scales for which the usual fluid description becomes inaccurate, or higher-order non-linearities become important, respectively.

\begin{figure}[htb]
\centering
$$
\includegraphics[width=0.5\textwidth]{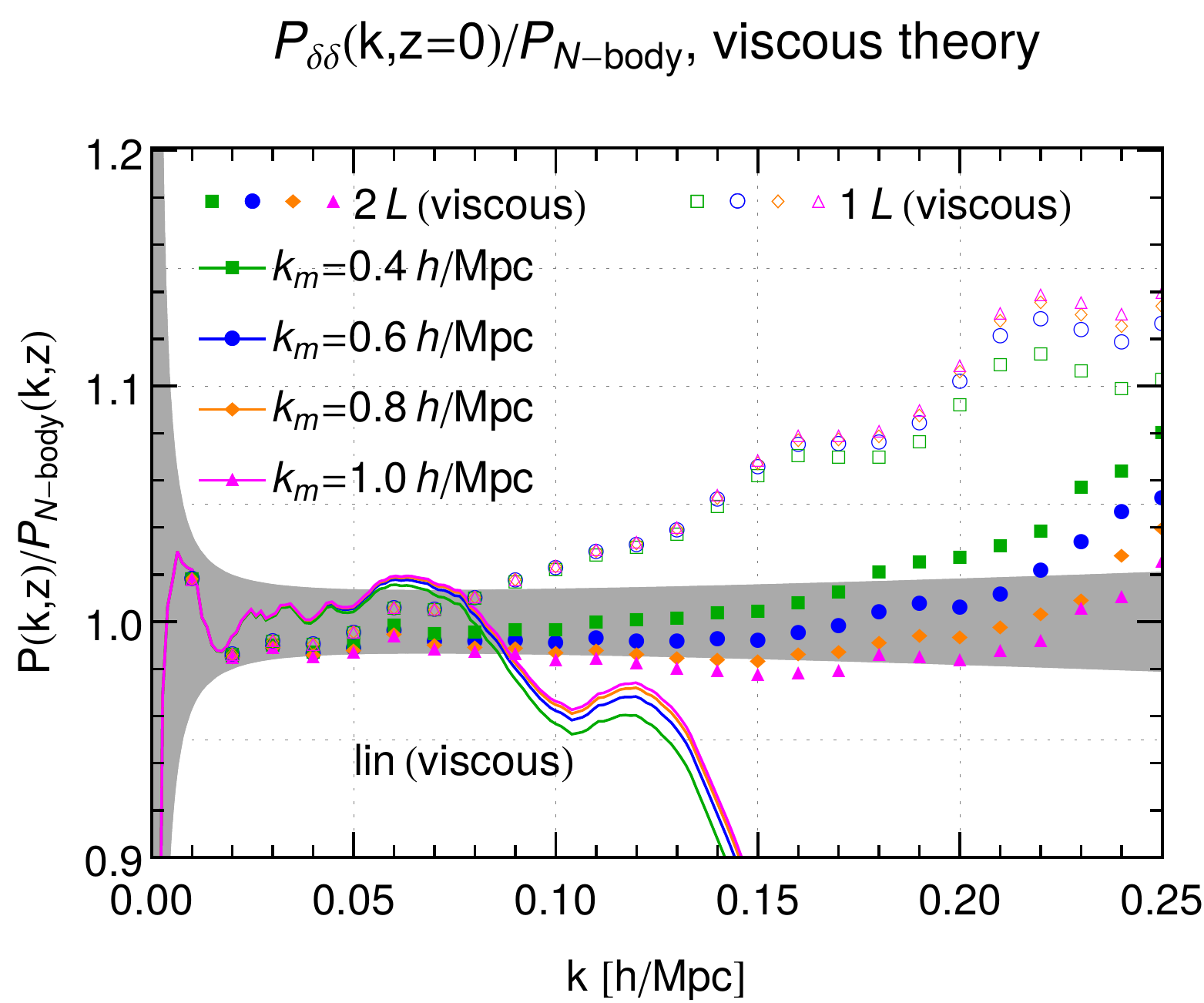}\qquad 
\includegraphics[width=0.5\textwidth]{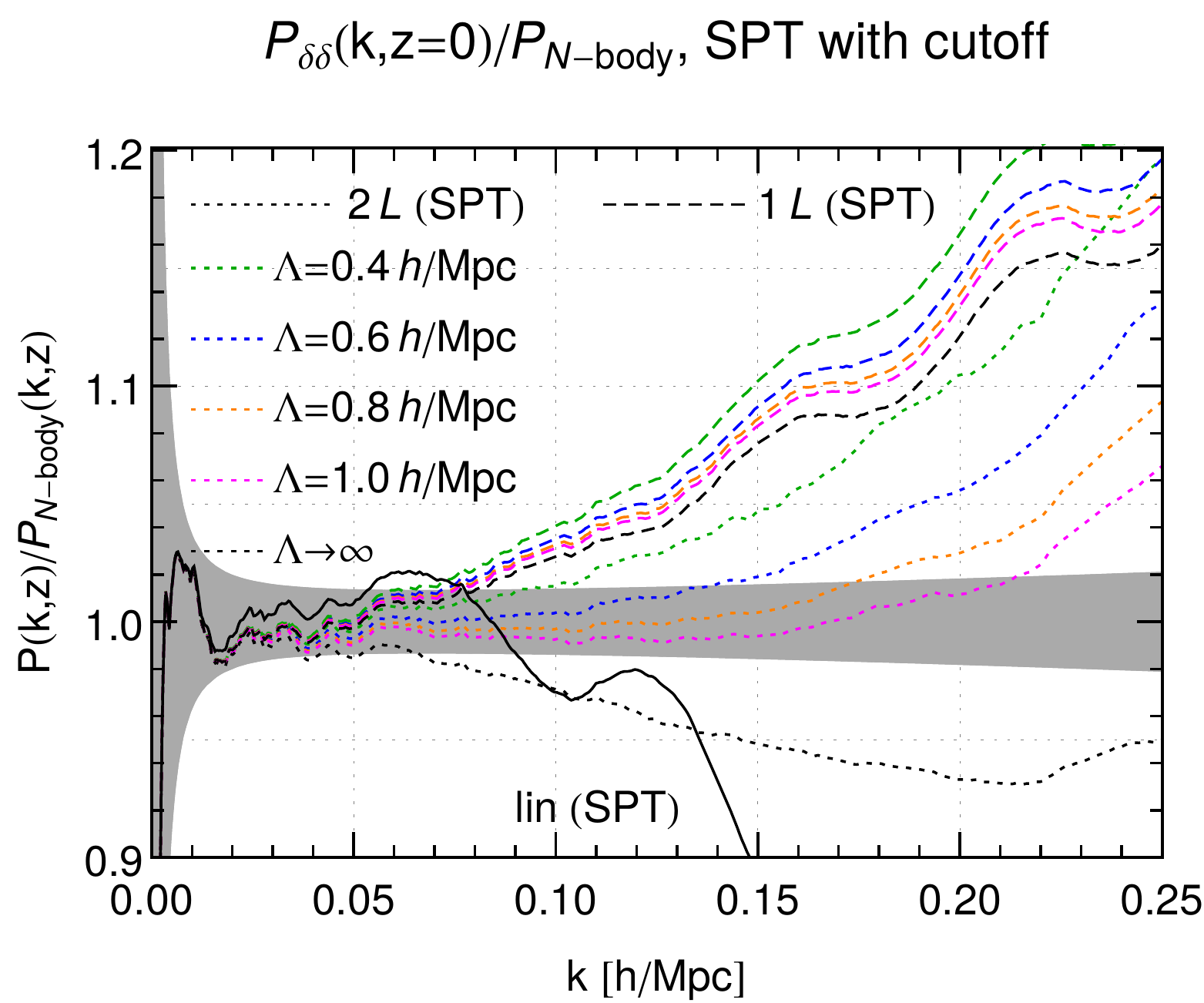}
$$
\caption{
Comparison of results for the power spectrum obtained within the viscous theory (left) and in standard perturbation theory (right)
at one- and two-loop level.
In this figure we normalized the power spectra to the $N$-body result~\cite{Kim:2011ab} at $z=0$. The grey band corresponds to
a $\pm 1\%$ deviation, plus an estimate for the error of the $N$-body results based on the boxsize $L=7200$Mpc$/h$ and
number of particles $N=6000^3$.
The left plot shows our one- and two-loop results within the viscous theory for various choices of the matching scale $k_m$, as in fig.\,\ref{figkm}. 
For comparison, the right plot shows
one- and two-loop results in standard perturbation theory (shown as dashed and dotted lines, respectively), computed with various
values of an ad-hoc cutoff $\Lambda$ (coloured lines),
as well as in the limit $\Lambda\to\infty$ (black lines).
}
\label{figkmnb}
\end{figure}

Importantly, also the size of the two-loop correction
relative to the one-loop is smaller by a factor of a few as compared to standard perturbation theory. This is partially
due to cutting off loop integrations at $k_m$, but can also be attributed to the reduced UV sensitivity within the viscous theory.
This can be observed in fig.\,\ref{figkmnb}, where we compare our results (left plot) in the viscous theory with standard
perturbation theory results (right plot). For the latter, we use an ad-hoc UV cutoff $\Lambda$ chosen to coincide with
the matching scale $k_m$. It is apparent from fig.\,\ref{figkmnb} that the two-loop result within standard perturbation theory
depends strongly on the cutoff $\Lambda$. In contrast, as discussed earlier, the one- and two-loop results within the viscous theory are significantly
less sensitive to the matching scale. We are aware of two effects that can contribute to a reduced UV sensitivity: first, as discussed
in subsection\,\ref{sec3.2}, the effective pressure and viscosity are dominated by modes close to $k_m$, i.e. the impact of very short scales on
modes with $k<k_m$ is actually small (in line with the findings of  \cite{Peebles:1980,Bagla:1996zb,nishimichi}, as discussed before). Second, the effect of pressure
and viscosity itself leads to a damping of UV modes, as discussed in detail in subsection\,\ref{sec2.2} for the linear solution. This damping
effect propagates also to the non-linear solution, in particular through the wavenumber-dependent terms in (\ref{omesplit}) appearing
on the left-hand side of (\ref{kernels}). For the second mechanism to be effective it is necessary to compute the kernels within
the viscous theory, rather than expand them in the pressure and viscosity contributions. The reason is that the latter strategy
would lead to contributions that grow with powers of the wavenumber, which artificially enhances the UV sensitivity, instead
of reducing it. On the other hand, the numerical treatment pursued here captures the damping effect which sets in
for modes close to $k_m$.

We conclude that the two-loop results are robust with respect to
a variation of the matching scale over the range $\sim 0.4-0.8\, h/$Mpc, and display an improved agreement
with $N$-body data as compared to the two-loop computed in standard perturbation theory.
In addition, our results suggest that the convergence of the loop expansion is improved compared to
the pressureless, ideal fluid case. This is an important property because it reduces the perturbative uncertainty.
 
\section{Conclusions}

We explored here a description of dark matter as a viscous fluid with non-vanishing sound velocity. For scalar perturbations in the regime of dark matter domination one finds that viscosity has the tendency to slow down gravitational collapse and structure formation in the linear regime for modes with large wavevector. However, in contrast to the effect of pressure, viscosity does not stop structure formation completely, and it does not wash out structures.

These properties are independent of whether viscosity and pressure have microscopic origin in terms of a non-trivial velocity distribution of the constituents of the dark matter fluid, or whether these terms arise on an effective level from the non-linear behavior of short wavelength modes.  It is this latter possibility that we explored in more detail.

More specifically, we obtained concrete expressions for the effective viscosity and sound velocity coming from short modes still within the fluid approximation by matching the linear propagator
of the viscous fluid to results for the propagator beyond linear approximation in standard perturbation
theory. The resulting values for viscosity and sound velocity depend on the cosmological time and on
the spectrum of inhomogeneities. Physically, these terms account for the leading effect that modes
with wavelengths below a coarse-graining scale $2\pi / k_m$ have on modes above that scale. 

For the modes with wavevectors $k<k_m$ we numerically solved the viscous fluid equations in linear, one-loop and two-loop approximation. The resulting density-density spectrum in the BAO range up to $k\lesssim 0.2 \, h/\text{Mpc}$ agrees on the percent level with the results from $N$-body simulations. 
This level of precision is consistent with the expected size of corrections from deep UV modes
for which the fluid approximation becomes inaccurate. The theoretical error can be estimated by the sensitivity to the matching scale $k_m$. We find that the two-loop
results within the viscous theory are robust at the percent level to a variation within the range $0.4-1\, h/$Mpc, in marked contrast to the cutoff dependence of the two-loop result within standard perturbation theory. Moreover, the effective pressure and viscosity lead to improved convergence of loop integrals in the ultraviolet so that one may hope for an improved convergence of perturbation theory also at higher orders.

In this sense, the theory with effective viscosity and pressure offers a convenient calculational framework for understanding inhomogeneities in the cosmological fluid at intermediate wavelengths. This set-up may be rather efficient in order to include also the leading effect of other phenomena that become relevant at short wavelength such as rotational (vector) degrees of freedom or heat transport due to interactions of the baryonic matter component. Finally, it would be intriguing to study the viscosity and pressure arising from kinetic motion and interactions in the dark matter sector in order to better understand its microscopic properties.


\end{document}